\documentclass[acmlarge, nonacm]{acmart}

\AtBeginDocument{  \providecommand\BibTeX{{    \normalfont B\kern-0.5em{\scshape i\kern-0.25em b}\kern-0.8em\TeX}}}

\usepackage{caption}
\usepackage{subcaption}
\usepackage{graphicx}  \graphicspath{{./}}
\usepackage{booktabs} \usepackage{bm}
\usepackage{mathtools}
\usepackage{listings}
\lstdefinestyle{BashInputStyle}{
	language=bash,
	basicstyle=\ttfamily,
	lineskip=-1pt,
	aboveskip=0pt, 	belowskip=0pt, 				frame=tb,
	framerule=0pt,
	columns=fullflexible,
	mathescape=true,
	escapechar=!
}
\usepackage{color}
\usepackage{tablefootnote}
\usepackage{accents}
\makeatletter
\newif\if@restonecol
\makeatother

\usepackage[lined,boxed,vlined,ruled]{algorithm2e}
\SetKwRepeat{Do}{do}{while}
\SetKw{Break}{break}
\SetKw{Return}{return}

\usepackage{amsthm}
\usepackage{amsmath}
\usepackage{amsfonts}
\usepackage{subcaption}
\usepackage{footmisc}
\usepackage{caption}
\usepackage{microtype}
\newcommand{\set}[1]{\{#1\}}

\newcommand{\defeq}{\stackrel{\text{def}}{=}}

\newcommand{\eat}[1]{}
\newcommand{\name}{\mbox{\texttt{HoneyComb}}\xspace}
\newcommand{\indexlayout}{\texttt{CoCo}\xspace}
\newcommand{\dom}{\textsf{Dom}}

\newif\iftechnicalreport
\technicalreportfalse

\newcommand{\revA}[1]{#1}
\newcommand{\revB}[1]{#1}
\newcommand{\revC}[1]{#1}

\newcommand{\avg}{\mathop{avg}}

\newcommand{\textsubscriptraise}[1]{\textsubscript{\raisebox{0.3ex}[-10pt][-10pt]{{\vphantom{i}#1}}}}
\newcommand{\tssr}[1]{\textsubscriptraise{#1}}

\begin{document}

\settopmatter{printacmref=false}
\setcopyright{none}
\renewcommand\footnotetextcopyrightpermission[1]{}
\pagestyle{plain}

\title{\name: A Parallel Worst-Case Optimal Join on Multicores}

\author{Jiacheng Wu}
\affiliation{  \institution{Univeristy of Washington}
  \city{Washington}
  \state{Seattle}
  \country{USA}}
\email{jcwu22@cs.washington.edu}

\author{Dan Suciu}
\affiliation{	\institution{Univeristy of Washington}
	\city{Washington}
	\state{Seattle}
	\country{USA}}
\email{suciu@cs.washington.edu}
\renewcommand{\shortauthors}{Jiacheng Wu and Dan Suciu}

\newcommand{\Perm}[1]{\ensuremath{{#1}_\sigma}}
\newcommand{\VarTuple}[1]{\ensuremath{\bm{#1}}}
\newcommand{\VarTuplePerm}[1]{\ensuremath{\bm{#1}_\sigma}}
\newcommand{\VarTupleProj}[2]{\ensuremath{\bm{#1}}_{#2}}
\newcommand{\VarTuplePermProj}[2]{\ensuremath{{\bm{#1}_{\sigma({#2})}}}}
\newcommand{\VTX}{\VarTuple{X}}
\newcommand{\VTXP}{\VarTuplePerm{X}}
\newcommand{\VTXPP}[1]{\VarTuplePermProj{X}{#1}}
\newcommand{\ConstTuple}[1]{\ensuremath{\bm{#1}}} 
\newcommand{\ConstTuplePerm}[1]{\ensuremath{\bm{#1}_\sigma}}
\newcommand{\ConstTupleProj}[2]{\ensuremath{\bm{#1}}_{#2}}
\newcommand{\ConstTuplePermProj}[2]{\ensuremath{{\bm{#1}_{\sigma({#2})}}}}
\newcommand{\ctx}{\ConstTuple{x}}
\newcommand{\ctxp}{\ConstTuplePerm{x}}
\newcommand{\ctxpp}[1]{\ConstTuplePermProj{x}{#1}}

\newcommand{\PartShare}{\ensuremath{\bm{P}}}
\newcommand{\PartSharePerm}{\ensuremath{\bm{P}_{\sigma}}}

\newcommand{\norm}[1]{\ensuremath{\left\lVert{#1}\right\rVert}}
\newcommand{\conj}[1]{\ensuremath{\bigwedge{#1}}}

\newcommand{\SetS}{\ensuremath{\mathcal{S}}}
\newcommand{\SetN}{\ensuremath{\mathcal{N}}}

\newcommand{\Concat}[2]{\ensuremath{#1 \oplus #2}}
\newcommand{\Restrict}[2]{\ensuremath{{#1}[{#2}]}}
\newcommand{\Cost}{\ensuremath{\mathcal{C}}}
\newcommand{\TotalCost}{\ensuremath{\mathbb{C}}}

\newcommand{\Optimal}[1]{\ensuremath{#1^*}}
\newcommand{\Skew}{\ensuremath{\gamma}}

\newcommand{\setlistingalphabet}{
  \captionsetup[lstlisting]{labelformat=simple, labelsep=colon}
  \renewcommand{\thelstlisting}{\Alph{lstlisting}}
}

\newcommand{\setlistingnumbering}{
  \captionsetup[lstlisting]{labelformat=simple, labelsep=colon}
  \renewcommand{\thelstlisting}{\arabic{lstlisting}}
}

\begin{abstract}
  To achieve true scalability on massive datasets, a modern query
  engine needs to be able to take advantage of large, shared-memory,
  multicore systems.  {\em Binary joins} are conceptually easy to
  parallelize on a multicore system; however, several applications
  require a different approach to query evaluation, using a Worst-Case
  Optimal Join (WCOJ) algorithm.  WCOJ is known to outperform
  traditional query plans for cyclic queries. However, there is no
  obvious adaptation of WCOJ to parallel architectures. The few
  existing systems that parallelize WCOJ do this by partitioning only
  the top variable of the WCOJ algorithm. This leads to work
  skew (since some relations end up being read entirely by every
  thread), possible contention between threads (when the
  hierarchical trie index is built lazily, which is the case on most
  recent WCOJ systems), and exacerbates the redundant computations
  already existing in WCOJ.

  We introduce \name, a parallel version of WCOJ, optimized for large
  multicore, shared-memory systems.  \name partitions the domains of
  all query variables, not just that of the top loop.  We adapt the
  partitioning idea from the HyperCube algorithm, developed by the
  theory community for computing multi-join queries on a massively
  parallel shared-nothing architecture, and introduce new methods for
  computing the shares, optimized for a shared-memory architecture.
  To avoid the contention created by the lazy construction of the
  trie-index, we introduce \indexlayout, a new and very simple index
  structure, which we build eagerly, by sorting the entire relation.
  Finally, in order to remove some of the redundant computations of
  WCOJ, we introduce a rewriting technique of the WCOJ plan that
  factors out some of these redundant computations. Our experimental
  evaluation compares \name with several recent implementations of
  WCOJ.
\end{abstract}


\maketitle

\section{Introduction} \label{sec:intro}

To achieve true scalability on massive datasets, a modern query engine
needs to be able to take advantage of large, shared-memory, multicore
systems.  Cloud companies offer servers with dozens of cores and many
terabytes of main memory; for example systems with up to 224 physical
cores and 24TB of main memory are available on
AWS~\cite{aws-ec2-high-memory}.  Users who need to perform complex
data analytics on large datasets will likely use such large servers
for their needs, and expect the query engine to scale up.  The
research community has studied parallel algorithms for query
processing extensively for over three decades, initially with a focus
on shared-nothing
architectures~\cite{DBLP:journals/tkde/DeWittGSBHR90,DBLP:journals/sigmod/DeWittG90,DBLP:journals/cacm/DeWittG92,DBLP:conf/vldb/DeWittNSS92},
but also on the shared-memory, multicore architectures that are the
focus of this
paper~\cite{DBLP:journals/vldb/BonczK99,DBLP:conf/sigmod/BlanasLP11,DBLP:conf/sigmod/ZhangHZH19,DBLP:conf/sigmod/ShahvaraniJ20,DBLP:conf/sigmod/0001MHGZHMM21,DBLP:conf/sigmod/0001MHGZHMM21,DBLP:conf/sigmod/WuWZ22}.
Many modern database engines use multi-threads during query
evaluation, and thus are prepared to use multiple cores, when they are
available.

{\em Binary joins} are conceptually easy to parallelize on a multicore
system: both relations are hash-partitioned, then the join is computed
in parallel in each partition.  Practical challenges consist of
reducing the number of blocking steps, reducing the number of cache
misses, and reducing the overhead of locks.  There exist several
solutions for these challenges, see for example the excellent
discussion in~\cite{DBLP:conf/sigmod/BlanasLP11}.

However, applications like graph databases, social network analysis, 
RDF/Sparql engines, queries on knowledge graphs, and even sparse tensor
compilers, are using a different approach to compute multi-join
queries, called \emph{Worst-Case Optimal Join},
WCOJ~\cite{DBLP:conf/icdt/Veldhuizen14,DBLP:conf/spire/BrisaboaCFN15,DBLP:conf/sigmod/ChuBS15,DBLP:conf/sigmod/AbergerTOR16,DBLP:journals/pvldb/FreitagBSKN20,DBLP:conf/sigmod/ArroyueloHNRRS21,DBLP:journals/pacmmod/WangWS23,DBLP:journals/sigmod/Salihoglu23,DBLP:conf/cidr/JinFCLS23}.
The WCOJ algorithms are theoretically optimal, and are known to
outperform binary joins on cyclic queries (but not on acyclic ones),
making them well-suited for these specialized engines and even for
some general-purpose relational
engines~\cite{DBLP:conf/sigmod/ArefCGKOPVW15,DBLP:journals/pvldb/FreitagBSKN20}.

Unfortunately, there is no obvious adaptation of a parallel 
hash-partitioned binary join to WCOJ.  Unlike a traditional query plan, a WCOJ
algorithm joins all relations at once.  The algorithm consists of
several nested loops, with one iteration for each join variable of the
query.  There exist a few parallel implementations of WCOJ, and they
parallelize only the topmost loop.  For example both
LogicBlox~\cite{DBLP:conf/sigmod/ArefCGKOPVW15} and
Umbra~\cite{DBLP:journals/pvldb/FreitagBSKN20} \revB{, as well as its Diamond-hardened extension~\cite{DBLP:journals/pvldb/BirlerKN24},} adopt this simple
parallelization strategy.  Under this approach, the system
hash-partitions the domain of the topmost variable into sets of equal
size, then executes the query for each partition in a separate logical
thread.  Each thread executes the remaining loops of the WCOJ
sequentially.  As a consequence, while each thread reads only a small
fragment of the relations that contain the first iteration variable,
it reads and processes the other relations entirely.  In other words,
relations that do not contain the top-most variable are not
partitioned at all under this approach.

\begin{figure*}[t]
  \setlistingalphabet
  \begin{minipage}[t]{.31\textwidth}

    \begin{lstlisting}[caption={Original Query}, style=BashInputStyle,
    lineskip=0pt,
    abovecaptionskip=0pt,
    label=lst:serial]
Q := R(X,Y), S(Y,Z), T(Z,X)

For x $\in$ !\colorbox{yellow}{R.X $\cap$ T.X}!
  For y $\in$ !\colorbox{green}{R[x].Y $\cap$ S.Y}!
    For z $\in$ !\colorbox{pink}{S[y].Z $\cap$ T[x].Z}!
      Q += (x, y, z)
\end{lstlisting}
   
  \end{minipage}
  \hfill
  \begin{minipage}[t]{.34\textwidth}
    \begin{lstlisting}[caption={Traditional Parallelization}, style=BashInputStyle,
    lineskip=-1pt,
    abovecaptionskip=0pt,
    label=lst:tradpara]  
// Parititon R.X and T.X on X
// In parallel,  Thread i:
For x $\in$ !\colorbox{yellow}{R\tssr{i}.X $\cap$ T\tssr{i}.X}!
  For y $\in$ !\colorbox{green}{R\tssr{i}[x].Y $\cap$ S.Y}!
    For z $\in$ !\colorbox{pink}{S[y].Z $\cap$ T\tssr{i}[x].Z}!
      Q += (x, y, z)
    \end{lstlisting}
    
  \end{minipage}
  \hfill
  \begin{minipage}[t]{.34\textwidth}
    \begin{lstlisting}[caption={\name's Parallelization}, style=BashInputStyle,
    lineskip=-1pt,
    abovecaptionskip=0pt,
    label=lst:newpara]  
// Partition R, S and T (see text)
// In parallel, Thread (i, j, k) 
For x $\in$ !\colorbox{yellow}{R\tssr{ij}.X $\cap$ T\tssr{ik}.X}!
  For y $\in$ !\colorbox{green}{R\tssr{ij}[x].Y $\cap$ S\tssr{jk}.Y}!
    For z $\in$ !\colorbox{pink}{S\tssr{jk}[y].Z $\cap$ T\tssr{ik}[x].Z}!
      Q += (x, y, z)
    \end{lstlisting}
    
  \end{minipage}

\caption{Comparison of WCOJ Parallelization Algorithms}
\label{fig:algo-comp}

\setlistingnumbering
\end{figure*}

In this paper, we introduce \name, a parallel version of WCOJ,
optimized for large multicore, shared-memory systems.  \name
partitions the domain of every variable, not just the top-most
variable, and creates a separate thread for every combination of
partitions.  This idea has been originally proposed for shared-nothing
architectures, first for MapReduce
systems~\cite{DBLP:conf/edbt/AfratiU10}, then was studied extensively
under a theoretical model of computation called Massively Parallel
Communication (MPC) model, where the algorithm is known as the {\em HyperCube Algorithm}~\cite{DBLP:journals/jacm/BeameKS17,DBLP:conf/icdt/KoutrisBS16,DBLP:journals/ftdb/KoutrisSS18,DBLP:journals/sigmod/HuY20,DBLP:conf/pods/Hu21}.

\revB{\name overcomes three limitations of the traditional WCOJ
  parallelization method: load skew, work duplication, and contention
  due to lazy index creation.  We start by illustrating the first
  limitation of the traditional method, load skew, using an example.}

\eat{
one
important limitation of the traditional parallelization approach:

the skew in the load of the threads.  We illustrate that with an
example.}

\begin{figure}[htbp]
  \centering
  \includegraphics[width=.75\linewidth]{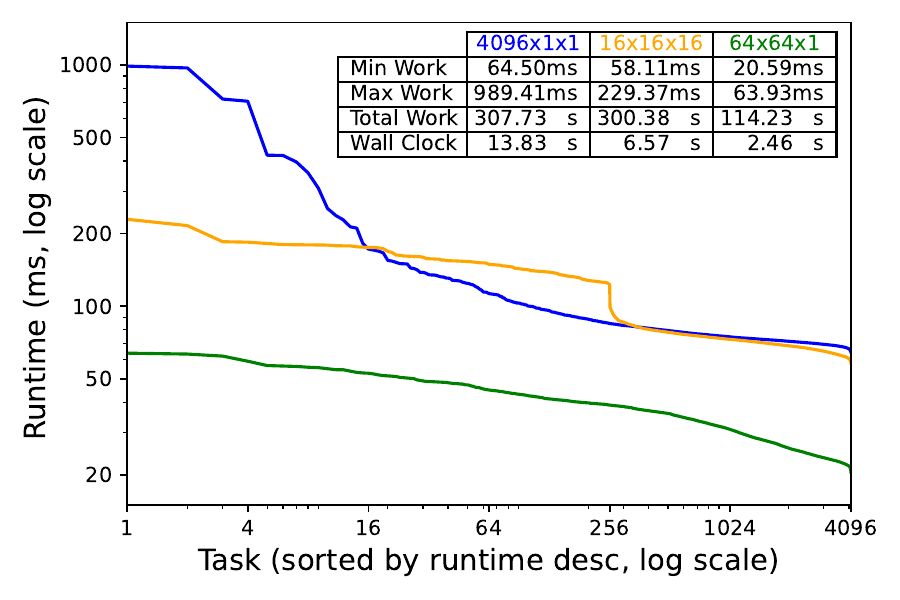}
  \caption[Skewness for Triangle Query on WGPB]{ \revB{Skewness for Triangle Query on WGPB. Here each of the three lines 
  
  represents the same number of partitions, $4096$, but allocated differently to the three attributes $X$, $Y$, and $Z$. The expression $4096 \times 1 \times 1$ (the blue line) means that the attribute $X$ was partitioned into $4096$ buckets and assigned to the threads, while the attributes $Y$, $Z$ were not partitioned (or partitioned into $1$ bucket). Similarly, $16 \times 16 \times 16$ (the orange line) means that each of $X$, $Y$, $Z$ was partitioned into $16$ buckets, for a total of $4096$ buckets/threads. Each point in the graph represents the runtime of one of the $4096$ threads, run in isolation, on one core; when executed on all $60$ cores, these threads will be dynamically scheduled on all $60$ cores, using work stealing.}}
  \label{fig:skew-eg}
\end{figure}

\begin{example} \label{ex:intro:1} Listing~\ref{lst:serial} in
  Fig.~\ref{fig:algo-comp} illustrates WCOJ with the triangle query
  $Q(X,Y,Z) = R(X,Y), S(Y,Z), T(Z,X)$.  We ran this query on the
  (undirected) WGPB~\cite{wgpb-dataset, DBLP:conf/semweb/HoganRRS19} dataset, with $54.0$M nodes and $81.4$M edges,
  using 4096 threads on a server with 60 physical cores and 120
  hyperthreads.\footnote{As we will see in Sec.~\ref{sec:exp}, \name keeps all cores busy, making hyperthreading ineffective, thus the ideal speedup is only a factor of 60, not 120.}  The traditional
  way to parallelize this query is shown in
  Listing~\ref{lst:tradpara}.  The domain of the first variable, $X$,
  is partitioned into 4096 fragments.  This partitions $R(X,Y)$ and
  $T(Z,X)$, but does not partition $S(Y,Z)$.  Each thread
  $i=1,2,\ldots,4096$ reads only a $1/4096$-fragment of $R$ and $T$,
  but reads the entire relation $S$.  We instrumented the code to
  measure, for each thread, its runtime:\footnote{In examples, we focus only on the join runtime,  given preprocessing is complete.} we show their distribution as
  the blue line in Fig.~\ref{fig:skew-eg}.  The skew (ratio between
  the largest and smallest runtime) is quite visible, showing a factor
  of more than 15. \revC{The cumulative time of the 4096 threads, i.e. the area under
  the curve, is 308s.  We used standard work-stealing to execute the
  threads on the 60 cores, thus, in an ideal scenario the total
  runtime should be $308/60 = 5.13$s: however, the actual measured
  total runtime (shown in the table at the top left) was 13.83s.
  As we will see, the discrepancy is mostly due to skew.}

\end{example}

Instead of the traditional parallelization technique, \name partitions
all query variables.

We illustrate this with an
example.

\begin{example} \label{ex:intro:2} Listing~\ref{lst:newpara} in
  Fig.~\ref{fig:algo-comp} shows how \name executes the triangle
  query, by partitioning each variable domain.  On the MPC model, the
  theoretically optimal partitioning is $16\times 16\times 16$,
  meaning that the domain of each variable $X, Y, Z$ is
  hash-partitioned into 16 buckets.  A thread is identified by a
  triple $(i,j,k)$, where each of $i,j,k$ range from $1$ to $16$, and
  it will only read a fragment of $1/16^2 = 1/256$ of each relation
  $R, S, T$.  The runtime distribution is the orange line in
  Fig.~\ref{fig:skew-eg}. \revC{The total work performed by the 4096
    threads (area under the curve) was $300$s, almost identical to the
    traditional method (Example~\ref{ex:intro:1}).  But} the skew
  ratio is only about 4, and the actual measured runtime on 60 cores
  decreased to 6.57s.  \revC{By reducing the load skew, \name reduced
    runtime by $2\times$.}

\end{example}

\revC{The second limitation of the traditional approach is work
  duplication.  For example, referring to Listing~\ref{lst:tradpara},
  the fragment $S[y]$ of the relation $S$ needs to be accessed by all
  4096 threads in order to perform the intersection
  $S[y].Z \cap T_i[x].Z$.  In order to reduce or to eliminate
  redundant work, \name departs from the theoretical MPC model,
  because the optimal partitioning strategy of the MPC model may also
  lead to work duplication. For example, in Listing~\ref{lst:newpara},
  the fragment $R_{ij}[x]$ of the relation $R$ needs to be accessed by
  all 16 threads $k=1,2,\ldots,16$ in order to perform the
  intersection $R_{ij}[x].Y \cap S_{jk}.Y$.}

\revB{One contribution in \name is a novel cost-based optimization of
  the number of buckets allocated to each variable, called the {\em
    share} of that variable.

  While the problem of computing optimal shares has been extensively
  studied by the MPC model, their solutions do not carry over to
  multicores.  The reason is that in the MPC model the communication
  is separated from the computation, and the optimal shares minimize
  only the communication cost, ignoring any work duplication at the
  servers.  In a shared-memory system there is no communication cost,
  instead the total work becomes the main cost.  \name computes the
  optimal shares using a cost model, which aims to minimize both skew
  and work duplication.}

\begin{example}
  The problem in the previous example is that, by partitioning the
  last variable $Z$, we create duplicated work for intersections of
  the previous variables.  Two threads with identifiers $(i,j,k_1)$
  and $(i,j,k_2)$ will see exactly the same partition $R_{ij}$.  In
  addition, if the data is dense, then they will likely also see
  similar values in $T_{ik}.X$, $S_{jk}.Y$. This causes portions of
  the top loops to be executed redundantly by different threads.  The
  choice of how many shares to allocate for each variable depends on
  the data distribution.  In our example, \name choose as optimal
  shares $64 \times64 \times 1$.  \revC{The load distribution for
    these shares is the green line in Fig.~\ref{fig:skew-eg}, and the
    cumulative load of all 4096 threads (area under the curve) has
    decreased to $114$s, from about $300$s for the
    $4096 \times 1 \times 1$ and the $16 \times 16 \times 16$ shares.
    As a consequence, the actual measured runtime on 60 cores
    decreased to $2.46$s.}

\end{example}

We introduced a novel cost model specifically for shared-memory,
parallel WCOJ, based on a pessimistic cardinality
estimator~\cite{DBLP:conf/sigmod/CaiBS19}, more precisely on a recent
refinement~\cite{DBLP:journals/pacmmod/KhamisNOS24} that uses more
sophisticated statistics on the input relations rather than just their
cardinalities.  To the best of our knowledge this is the first cost
model proposed for WCOJ.\footnote{Recent
  work~\cite{DBLP:journals/pvldb/WangT0O23} circumvents the need for a
  cost model for WCOJ by using Reinforcement Learning for choosing the
  variable order.}  Our cost model is described in
Sec.~\ref{subsec:cost:model}.

\revB{Finally, the third limitation of the traditional approach comes
  a particular optimization in modern WCOJ systems, the lazy
  construction of the hash-trie, which in the parallel setting leads
  to contention between threads.}

All WCOJ implementations index the data, in order to speed up the
intersections that need to be done at each iteration of the algorithm.
While a standard trie was used in an early implementation of
WCOJ~\cite{DBLP:conf/icdt/Veldhuizen14}, recent implementations use a
hash trie index.  For each relation, the first level of the trie is a
hash table consisting of values of the first variable, pointing to
hash tables consisting of values of the second variable, and so on.
Since the hash trie construction is expensive, recent systems compute
lazily the entries on levels two and higher: only sub-tries that are
actually needed are
constructed~\cite{DBLP:journals/pvldb/FreitagBSKN20,DBLP:journals/pacmmod/WangWS23}.
We illustrate this in Fig.~\ref{fig:hashtrie}, where the lazy sub-trie
construction applies to $R.Z$ and $T.Z$.  However, when done in
parallel, the lazy index requires all index accesses to be coordinated
via locks, in order to avoid read-write conflicts.  This is not a
problem for a small number of cores, say 12-16, but it becomes a
problem for a larger number of cores, which is our target for \name.

To avoid this contention and further remove the need for locks, \name
introduces a new, and simple sort-based trie index that we call
Compressed Column layout, or \indexlayout.  Level $k$ of \indexlayout
is a sorted array, consisting of the first $k$ attributes of the
relation, which pointers to the next level of the trie.  We construct
\indexlayout eagerly, by first sorting the entire relation once, then
computing each level by a projection and duplication elimination.  We
benefit from the fact that there exists very effective parallel,
shared-memory sorting algorithm: \name uses \textbf{ips4o} (In-place
Parallel Super Scalar Samplesort)~\cite{axtmann2017,
  axtmann2020engineering} for this purpose.  By replacing the
hash-based index with a sort-based one we reduce the construction time
significantly, and this allows us to compute it eagerly, and avoiding
any need for coordination during the parallel execution of
WCOJ. \revB{Thus, \indexlayout represents a return to the original
  trie in LFTJ~\cite{DBLP:conf/icdt/Veldhuizen14}, but in a simplified
  form, where the entire level $k$ is a single array, which allows us
  to construct the entire index using a single parallel sort.}

Finally, a last challenge of the WCOJ algorithm that we observed in
\name is that it some duplicated work can be avoided by storing
intermediate results and using them later.  Each loop of a WCOJ
computes the intersection of all domains of the current variable.
Some of these intersections are computed repeatedly, for different
iterations of some outer loop.  This occurs in both sequential and
parallel implementations of WCOJ, but it seems to impact more the
parallel implementation.  To address that, we introduce an
optimization that we call a \emph{rewriting mechanism}.  We precompute
this intersection during the outer loop, then take advantage of the
shared memory that allows multiple threads to read that intermediate
result.  We describe the rewriting mechanism in
Sec.~\ref{subsec:rewrite}.

We have evaluated our methods and compared them with the state-of-the-art
parallel implementations of WCOJ. Our approach, \name, demonstrates
strong scalability, achieving near-linear scaling observed up to 60 cores.
\name outperforms other baselines by 3x on the WGPB dataset, which
consists of sparse and skewed data, and can be 10x to 100x faster than
competing systems on dense data.  We report our evaluation in
Sec.~\ref{sec:exp}.

In summary, we make the following contributions: 

\begin{itemize}
\item We describe \name, a parallel version of WCOJ that adapts the
  HyperCube algorithm to a multicore system (Sec.~\ref{sec:join}).
\item We present a novel trie structure index \indexlayout that faciliates and accelerates the operations of WCOJ (Sec.~\ref{sec:prep}).
\item We introduce a rewriting mechanism that further reduces redundant
  computation of the parallel WCOJ (Sec.~\ref{subsec:rewrite}).
\item We propose a new cost model for the parallel WCOJ and use it for
  computing an optimal partition share allocation (Sec.~\ref{subsec:cost:model} and Sec.~\ref{subsec:plan}).
\item We conduct an extensive experimental evaluation
  (Sec.~\ref{sec:exp}).
\end{itemize}

\section{Background and Problem Statement} \label{sec:bg}

{
\begin{table}[t]
	\centering
	\caption{Basic Notations for Cost Model}
	\label{tab:freq}
	\begin{tabular}{cc}
          \toprule
          Notations & Definition \\
          \midrule
          $\ctx$ / $\VTX$ & tuple of constants / variables \\
          $\ctxp$ / $\VTXP$ & permuted tuple of constants /
                              variables \\
          \ \ \ $\ConstTupleProj{x}{i:j}$  /
          $\VarTupleProj{X}{i:j}$
                    & projected tuple of constants /
                      variables\tablefootnote{$\ctxp$ means
                      $(\VarTupleProj{x}{\sigma(1)},\VarTupleProj{x}{\sigma(2)},\ldots\VarTupleProj{x}{\sigma(n)})$
                      and $\VarTupleProj{x}{i:j}$
                      means $(\VarTupleProj{x}{i},\VarTupleProj{x}{i+1},\ldots\VarTupleProj{x}{j})$.} \\
          $\Concat{\ctx}{\VTX}$ & concatenation of constants or variables\\
          $\SetS$ & set of relations, e.g. $\{R_{j_1}, \cdots, R_{j_k}\}$\\
          $|\SetS|$ & the size of $\SetS$, e.g. $k$ above \\

          $\SetN$ &  cardinalities of relations in $\SetS$, e.g. $\set{|R_{j_1}|,\ldots,|R_{j_k}|}$ \\
		\bottomrule 
\end{tabular}
\end{table}
}
{
  \setlength{\abovedisplayskip}{0.25em}  
  \setlength{\belowdisplayskip}{0.25em} 
\noindent A \textbf{full Conjunctive Query} with variables $X_1,
\ldots, X_n$ is defined as:
\begin{equation}
  Q(X_1, \ldots, X_n) \leftarrow R_1(\bm{X}_1)\wedge R_2(\bm{X}_2)\wedge \dots\wedge R_m(\bm{X}_m)\label{eq:full:cq}
\end{equation}
Each term \( R(\bm{X}_i) \) is called an \emph{atom}, where
\( \bm{X}_i \subseteq \set{X_1, \ldots, X_n}\) is a tuple of
variables, and every variables $X_i$ occurs in some atom
$R_j(\bm X_j)$. }We use $m$ to denote the number of atoms, and $n$ for
the number of variables; we also use normal letters $X_i$ to denote
single variables, and boldface $\bm X_j$ to denote a tuple of
variables.

We will always drop the head variables, since they are clear from the
rest of the query, for example we abbreviate a 2-way join
$Q(X,Y,Z)=R(X,Y)\wedge S(Y,Z)$ as $Q = R(X,Y)\wedge S(Y,Z)$.
Predicates are pre-processed by pushing them down to the atoms, for
example we treat the query $Q=R(X,Y)\wedge S(Y,Z) \wedge Z>5$ as
$Q = R(X,Y)\wedge S_0(Y,Z)$ where $S_0=\sigma_{Z>5}(S)$.  We do not
discuss predicates in this paper.

\addvspace{\smallskipamount}
\noindent\textbf{Generic Join Algorithm} Traditional query engines compute the
query~\eqref{eq:full:cq} one join at a time.  The problem with this
approach is that the size of intermediate relations can be larger than
the final output.  For example, if each of the three relations
$R, S, T$ of the triangle query in Fig.~\ref{fig:algo-comp} has
size $N$, then the final output is $\leq N^{3/2}$, while any 2-way
join can have size $N^2$.  \emph{Generic Join}, GJ, which is a special
case of a \emph{Worst Case Optimal Join}
(WCOJ)~\cite{DBLP:journals/sigmod/NgoRR13}, computes the query in time
that is guaranteed to be no larger than the largest possible output to
$Q$.  GJ chooses some variables order, say $X_1, X_2, \ldots$, then
performs $n$ nested loops, one for each variable $X_i$. At each loop
it first computes the intersection of all columns $R_j.X_i$, the binds
the variable $X_i$ to each value $x_i$ in the intersection and
computes the residual query, where each $R_j$ containing $X_i$ is
restricted to $R_j[x_i]$:

\centerline{
  \begin{minipage}{0.7\linewidth}
  \begin{tabbing}
    \texttt{for} \=$x_1$ \texttt{in} $R_{j_1}.X_1 \cap R_{j_2}.X_1 \cap \ldots$ \\
    \>\texttt{for} \=$x_2$ \texttt{in} $\ldots$\\
    \> \>\texttt{for} \=$x_3$ \texttt{in} $\ldots$\\
    \> \> \>$\ldots$
  \end{tabbing}
\end{minipage}
}

Referring to the triangle query in Fig.~\ref{fig:algo-comp}, the
top loop binds $X$ to each value of the intersection $R.X\cap T.X$,
then restricts $R, T$ to $R[x], T[x]$ (their subsets where $X=x$) and
computes the residual query $R[x](Y) \wedge S(Y,Z) \wedge T[x](Z)$.
The runtime of GJ is proven to be no larger than the maximum output
size.  The theoretical guarantee holds for any variable order, but, in
practice, the choice of variable order can affect the runtime
significantly~\cite{DBLP:journals/pvldb/WangT0O23}.

\begin{figure}[t]
	\includegraphics[width=.9\linewidth]{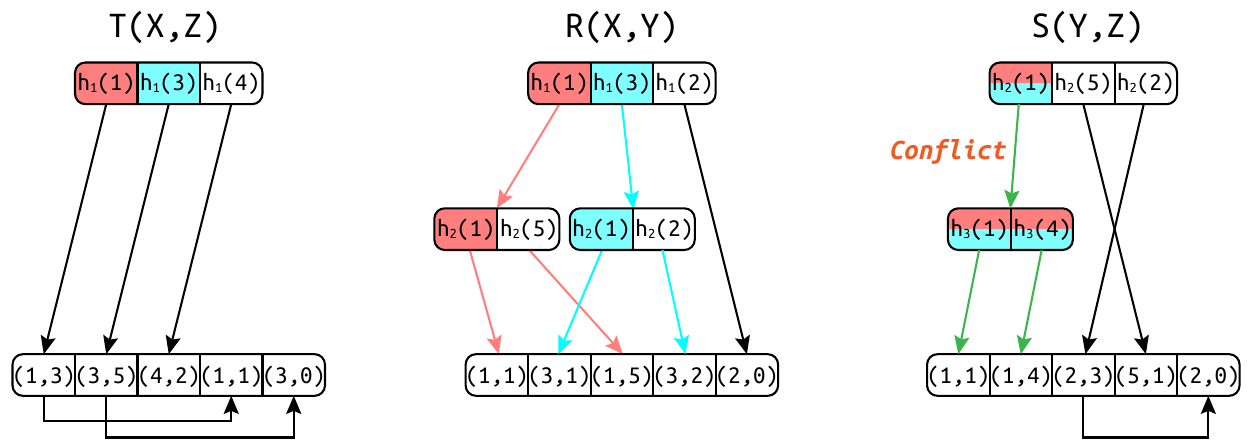}
	
	\caption{(Adapted from~\cite[Fig.6]{umbratechreport})
          Conflicts in Traditional Parallel WCOJ Algorithm. The blue
          thread tries to access the sub-trie $h_2(1)$ in $S.Y$, while
          the red thread is constructing it. }
  \label{fig:hashtrie}
\end{figure}

An important aspect of GJ is that, in order to guarantee the
theoretical runtime, it must compute the intersection
$R_{j_1}.X_i \cap R_{j_2}.X_i\cap \cdots$ in time proportional to the
smallest of the sets.  This is achieved by representing each relation
as a hash trie with a depth equal to the number of its attributes,
where each level corresponds one attribute, see
Fig.~\ref{fig:hashtrie}.  To compute the intersection, GJ selects the
smallest hash table, iterates over the values stored there, and probes
each of them in all the other hash tables.  Since pre-computing all
hash tries for all relations are expensive, a common technique is to
use a \emph{lazy trie}, where the trie is constructed on demand, as
needed during the
join~\cite{DBLP:journals/pvldb/FreitagBSKN20,DBLP:journals/pacmmod/WangWS23}.
Thus, only relevant parts of the trie are built.

\addvspace{\smallskipamount}
\noindent\textbf{Parallel GJ} Logicblox\cite{DBLP:conf/sigmod/ArefCGKOPVW15}
and Umbra~\cite{DBLP:journals/pvldb/FreitagBSKN20} parallelilze GJ by
partitioning the domain of the first variable $X_1$ into subsets of
equal size, then computing the query on each subset in a separate
thread, see Listing~\ref{lst:tradpara} in Fig.~\ref{fig:algo-comp}.  The domains of
$X_2, X_3, \ldots$ are not partitioned, and they will be scanned
entirely by every thread, which leads to two important drawbacks.
First, even if the domain of $X_1$ is uniformly distributed, the total
work may be skewed: this can be seen for example in
Fig.~\ref{fig:skew-eg}.  Second, as shown in Fig.~\ref{fig:hashtrie}, since multiple threads access the
same values of $X_2, X_3, \ldots$ they may attempt to construct
concurrently the same fragment of the lazy trie, requiring the use of
locks for synchronization (to avoid read/write conflicts), which
introduces significant overhead. Those conflicting overheads are revealed by the scalability experiments of Umbra in Sec.~\ref{subsec:exp:scale}.

\begin{figure}[t]
	\includegraphics[width=.65\linewidth,trim={0 0 0 0},clip]{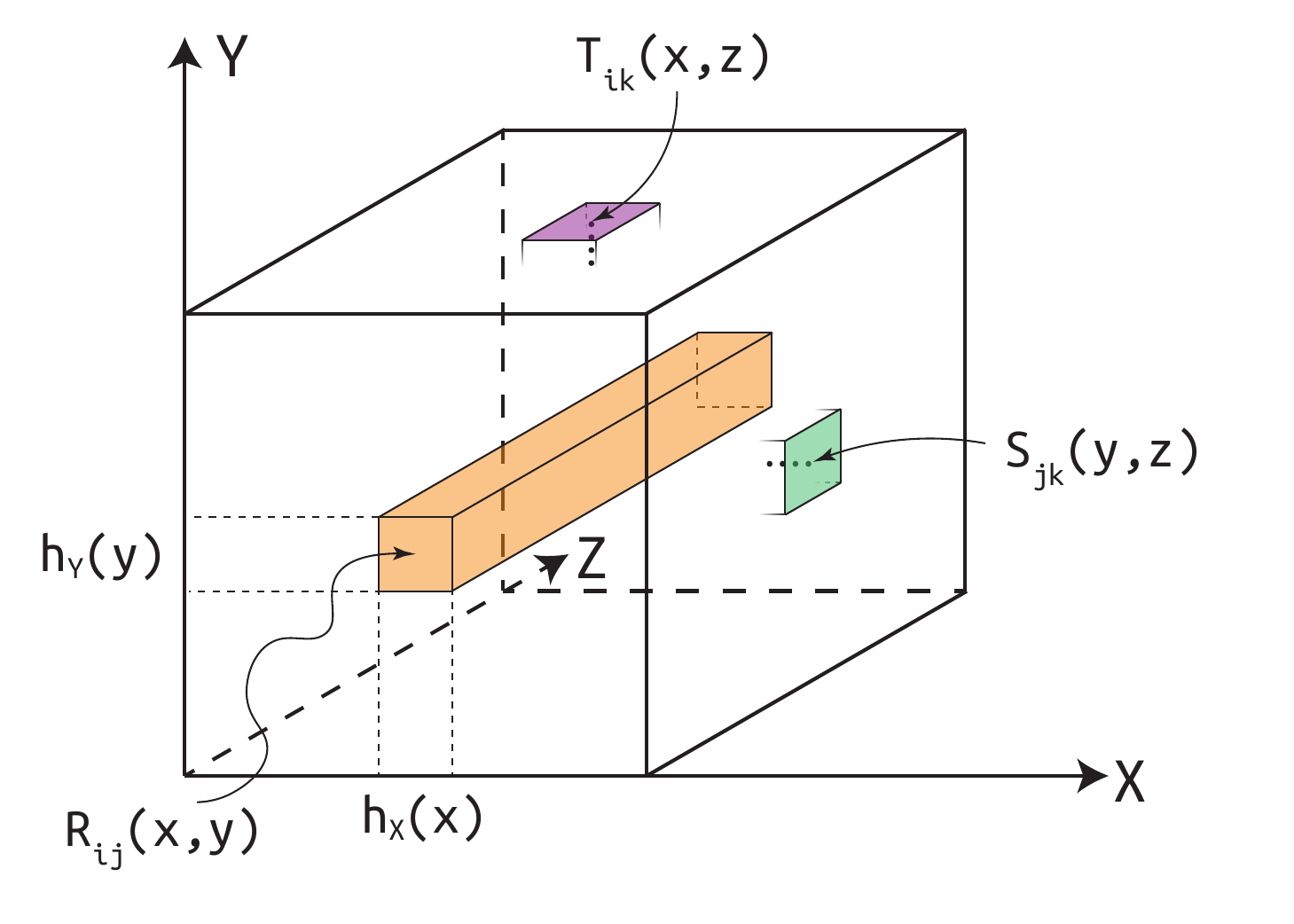}
	\Description[HyperCube]{HyperCube Partition}
	\caption{\revA{The HyperCube Partition}}
  \label{fig:hc}
\end{figure}

\addvspace{\smallskipamount}
\noindent\textbf{HyperCube Algorithm:} The HyperCube algorithm~\cite{DBLP:conf/edbt/AfratiU10,DBLP:journals/jacm/BeameKS17,DBLP:conf/icdt/KoutrisBS16}
computes the query~\eqref{eq:full:cq} on a shared-nothing architecture
with $P$ servers by partitioning \emph{all} domains of \emph{all}
attributes. It organizes the $P$ servers into a hypercube with $n$
dimensions, by writing $P$ as a product
$P = P_1 \times P_2 \times P_n$, such that each server is uniquely
identified by $n$ coordinates, $(s_1, \ldots, s_n)$, with
$s_i \in \set{0,1,\ldots,P_i-1}$.  In the first step, HyperCube hash-partitions the
domain of each variable $X_i$ into $P_i$ buckets then, in a single
global communication step, it sends every tuple $R_j(\ldots)$ to all
servers whose coordinates agree with the hash values of the tuple. In
the second step, each of the $P$ servers computes the query on the
fragment of relations it has received.  The quantity $P_i$ is called
the \emph{share} of the variable $X_i$.

For example, consider the query $Q=R(X,Y),S(Y,Z),T(Z,X)$ and $P=4096$
severs.  Hypercube writes $P=16 \times 16 \times 16$ and assigns to each
server a unique combination of 3 coordinates
$(s_1, s_2, s_3) \in
\set{0,\ldots,15}\times\set{0,\ldots,15}\times\set{0,\ldots,15}$, \revA{see
  Fig.~\ref{fig:hc}.} 
In the first step, it sends each tuple $R(x,y)$ to all of servers with the
coordinates\footnote{HyperCube uses independent hash functions
  $h_X, h_Y, h_Z$ for each coordinate.}  $(h_X(x),h_Y(y), *)$; each
tuple $(y,z)$ in the relation $S$ is also sent to all servers $(*,h_Y(y),h_Z(z))$, and
similarly for $T$. In the second step, each server computes the query
on the relation fragments that it has received.  Notice that each
relation is replicated 16 times, for example each tuple $R(x,y)$ is
sent to 16 servers, $(s_1,s_2,0), \ldots, (s_1,s_2,15)$.

A direct adaptation of HyperCube from the shared-nothing to a
multicore, shared-memory architecture will have poor performance.
The theoretical analysis in~\cite{DBLP:journals/sigmod/KoutrisS16}
addresses \emph{only} the communication cost, i.e. the total number of
tuples received by any server, and optimizes the shares
$P_1, P_2, \ldots, P_n$ such as to minimize the communication cost.
No algorithm is prescribed for the second step, and its computational
cost is not analyzed.  As we show in our paper, for multicores the
choice of the variable order and the associated shares must be
considered together.

\addvspace{\smallskipamount}
\noindent\textbf{Sparse Matrix Format}:

The high-performance, and the sparse tensor compiler communities have
developed as a suite of very efficient main memory representations for
sparse tensors.  Examples include Compressed Sparse Row (CSR),
Compressed Sparse Column (CSC),
Compressed Sparse Block (CSB)
the Coordinate List (COO), Blocked
Compressed Sparse Row (BCSR), Double Compressed Sparse Row (DCSR),
see~\cite{DBLP:conf/spaa/BulucFFGL09},
\cite[Fig.5]{DBLP:journals/pacmpl/KjolstadKCLA17} and
\cite[Fig.2]{DBLP:journals/pacmpl/ChouKA18}.  Our
representation~\indexlayout adapts ideas from sparse tensor
representations for parallel evaluation of Generic Join.

\addvspace{\baselineskip}

\noindent\textbf{Problem Statement} The goal of this paper is to
design efficient and scalable parallel variants of Generic Join; we
will use the terms Generic Join and Worst Case Optimal Join
interchangeably.  As we saw, the traditional approaches to
parallelization has several shortcomings. It is sensitive to
\textbf{data skew}, causing some processors to be overloaded while
others remain underutilized, thus degrading parallelization
efficiency.  The lazy trie construction \textbf{read-write conflicts}
when executed in multi-threaded environments, which requires the use
of locks for each index access.  A third shortcoming (described in
Sec.~\ref{sec:opt}) is that some duplicated work that is inherent in
Generic Join impacts the runtime, and this is more apparent in
parallel implementations.

Our paper proposes, \name, designed to address these problems by
ensuring better workload balance and avoiding conflicts during
parallel execution.  Additionally, \name is optimized for modern
hardware architectures, by minimizing pointer chasing to improve
memory access patterns and leveraging vector processing capabilities
to accelerate operations.

\section{Architecture of \name}

\begin{figure}[t]
	\includegraphics[width=.75\linewidth]{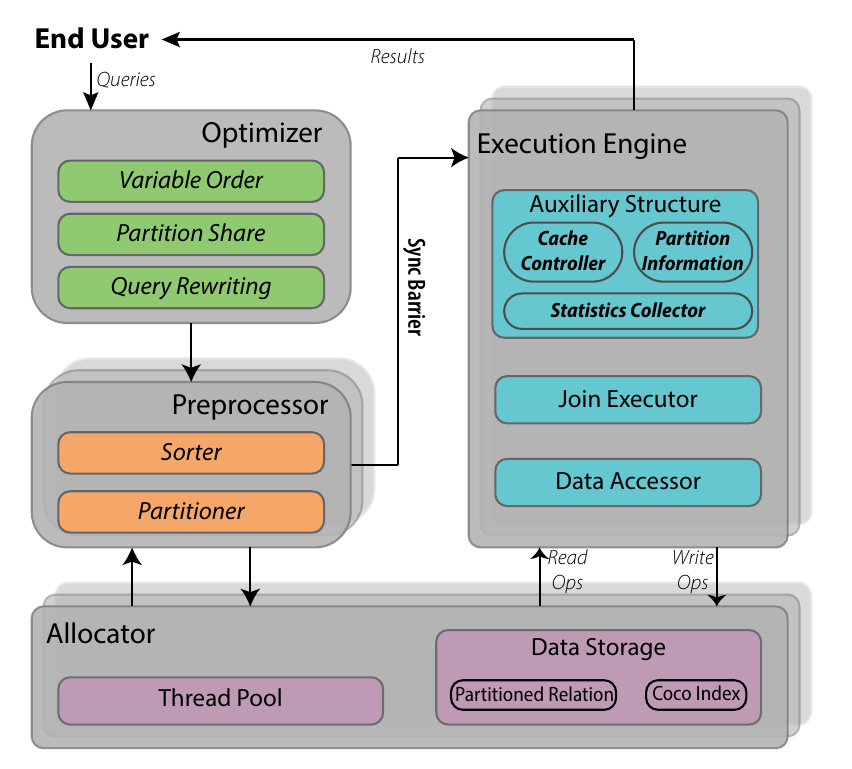}
	\Description[Arch]{The Architecture of \name}
	\caption{The Architecture of \name} 
	\label{fig:arch}
\end{figure}

\name takes as input a conjunctive query (see Eq.~\eqref{eq:full:cq})
and evaluates it in parallel, using a fixed number of threads $P$; we
use by default $P=1024$ threads.  The threads are executed in parallel
by all available cores on the system.  Since \name assumes that the
system has a large number of cores, e.g.  dozens, it aims to avoid any
synchronization between threads, because these can lead to significant
slowdown when the number of cores is large.  

\name uses the same
partitioning method as HyperCube, by computing a number of shares
$P_i$ for each query variable $X_i$, such that
$P_1 \times P_2 \times P_n=P$, but, unlike HyperCube, it does not
replicate the data, instead leverages the shared memory and allows
threads to read concurrently.  It optimizes both the variable order
and the shares together.  To avoid read-write conflicts, it
precomputes all trie indices eagerly, using a novel index
called~\indexlayout, which is optimized for modern hardware
architecture by minimizing pointer chasing and leveraging vector
processing capabilities.  

\name has four parts, see Fig.~\ref{fig:arch}: the Optimizer, the
Preprocessor, the Executor, and the Allocator.

The \textbf{optimizer}, described in Sec.~\ref{sec:opt}, uses a cost
model (Sec.~\ref{subsec:cost:model}) to determine both the variable
order and the shares for each variable (Sec.~\ref{subsec:plan}).  The
optimizer also performs query rewriting to remove some redundant
computations, described in Sec.~\ref{subsec:rewrite}.

Once the shares for each variable are computed, the
\textbf{Preprocessor} (Sec.~\ref{sec:prep}) sorts the input relations,
physically partitions them, and builds the~\indexlayout index for each
partition.  The partitioner is fully parallelized and its threads do
not require coordination.

Finally, the \textbf{Executor} (Sec.~\ref{sec:join}) performs the
actual work, by computing the query on each partition.  The algorithm
is similar to the standard Generic Join, with two exceptions: it uses
our sort-based index~\indexlayout instead of a hash-trie, and it may
compute and store some intermediate results, as introduced by the
optimizer during source rewriting. The executor is fully parallelized,
and its threads do not require coordination; as we show in
Sec.~\ref{sec:exp}, this allows it to scale almost linearly up to 60
physical cores.

The \textbf{Allocator} uses use Intel Thread Building Block (TBB) to
implement the scheduling framework~\cite{inteltbb}.  This is a
standard task-stealing technique to execute threads on the available
cores.  The framework dynamically allocates computational tasks to
cores, ensuring that all cores are engaged in meaningful work
throughout the execution process. Moreover, the Allocator is designed
to provide isolated partitioned data and cache for both the Executor
and Preprocessor, minimizing contention and maximizing efficiency.
Since this is a standard technique, we will not discuss the allocator
in this paper any further.

\section{The Preprocessor} \label{sec:prep}

The preprocessor is responsible for partitioning the input relations,
and for computing the trie indices, called \indexlayout, for each
partition.  It receives the shares $P_1, P_2, \ldots, P_n$, and the
variable order $X_{\sigma(1)}, \ldots, X_{\sigma(n)}$ from the optimizer; for presentation
purposes we assume in this and the next section that the variable
order is $X_1, X_2, \ldots, X_n$, and we will revisit this in
Sec.~\ref{sec:opt}.

\begin{figure}[t]
  \centering
  \includegraphics[height=.33\textheight]{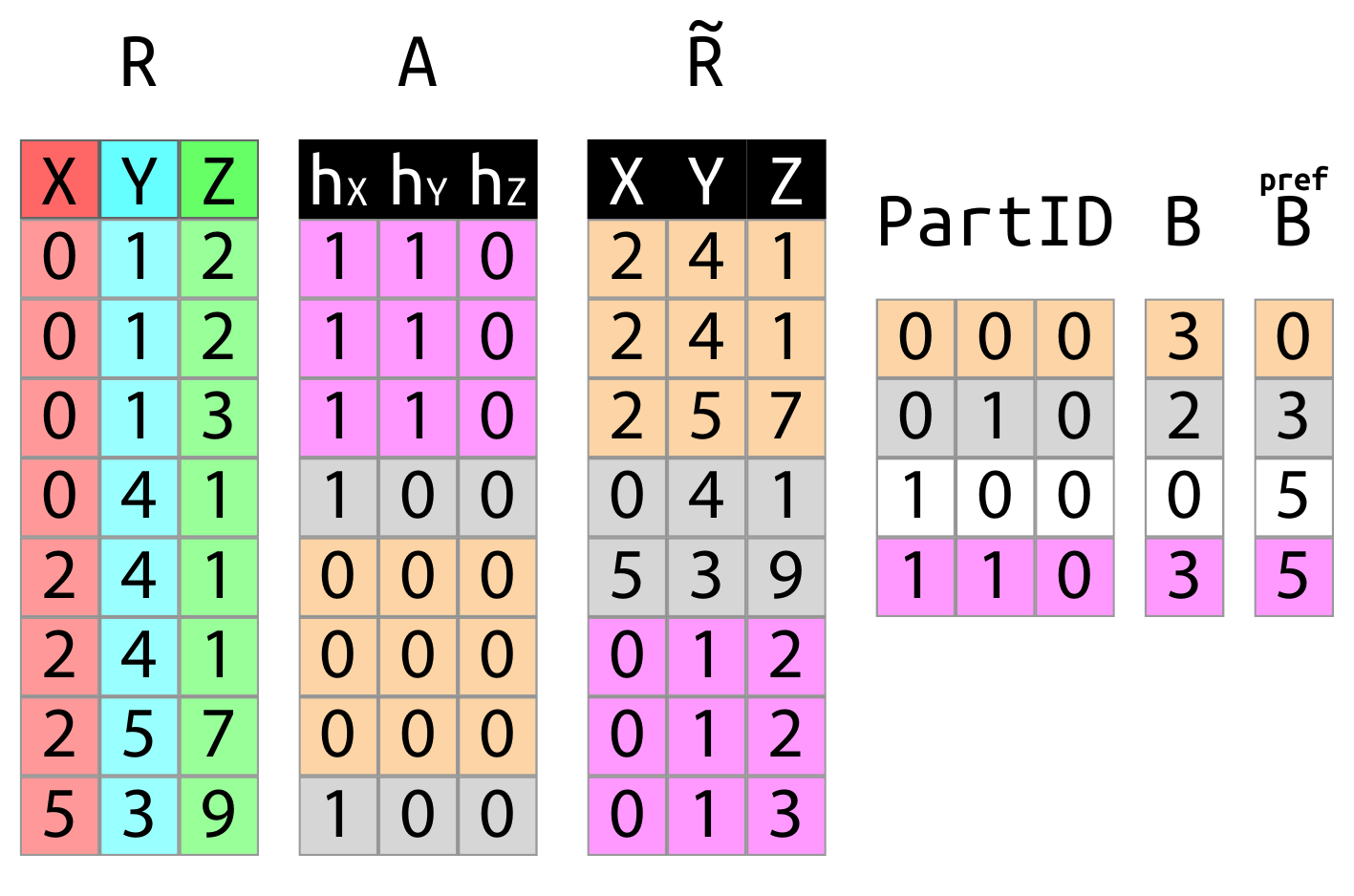}
  \caption{Parallel Partitioning}
  \label{fig:parapart}
\end{figure}

The input data is partitioned using the HyperCube method.  The
preprocessor chooses $n$ independent hash functions
$h_1, \ldots, h_n$, one for each query variable.  We will assume
w.l.o.g. that $h_i$ returns numbers in the set
$\set{0,1,\ldots,P_i-1}$ (otherwise we replace it with
$h_i(x) \mod P_i$).  Initially, each input relation
$R_j(X_{i_1},\ldots,X_{i_k})$ is stored in an array; we follow the C
convention and assume array indices start at 0.  Consider a $k$-tuple
$(x_1, \ldots, x_k)$ of $R_j$.  The quantities
$s_1 \defeq h_{i_1}(x_1), \ldots, s_k \defeq h_{i_k}(x_k)$ are called
the {\em partition identifiers} of this tuple.  The preprocessor
physically partitions $R_j$ into
$\Pi_j \defeq P_{i_1} \! \times \cdots \times  P_{i_k} \leq P$ chunks, called partitions,
where each partition $R_{j,s_1,\ldots,s_k}$ is a continuous array
holding all tuples with partition identifiers $s_1, \ldots, s_k$; all
partitions are concatenated and stored in an output array $\tilde R_j$
of the same size as $R_j$.  Unlike HyperCube, in \name there is no
need to replicate $R_j$, instead, the partitioned data $\tilde R_j$ is
an array of exactly the same size as $R_j$.  As we discuss in the next
section, during execution multiple threads will read from the same
partition, taking advantage of the shared memory.

We describe now the details of the parallel partitioning step.  First,
\name computes (in parallel) the partition identifiers
$(s_1,\ldots, s_k)$ of each tuple $(x_1, \ldots, x_k)$ in $R_j$, by
applying the hash functions, and storing them in an array $A_j$, of the
same size as $R_j$.  

Next, it computes the size of each partition
$R_{j,s_1,\ldots,s_k}$ and stores them in a $k$-dimensional array
$B_j$.  Next, it computes a prefix sum on $B_j$ and stores the result
in $B^{pref}_j$: notice that both $B_j$ and $B^{pref}_j$ have size $\Pi_j$ (the
number of partitions of $R_j$), which is $\leq P$ (the total number of
threads).  $B_j$ is computed by scanning the tuples
$(s_1, \ldots, s_k)$ in $A_j$ and incrementing
$B_j[s_1, \ldots, s_k]$; while $B^{pref}_j$ is the prefix sum of $B_j$,
$B^{pref}_j[s_1, \ldots, s_k] = \sum_{(t_1,\ldots,t_k)\preceq
  (s_1,\ldots,s_k)} B_j[t_1,\ldots,t_k]$, where $\preceq$ represents
lexicographic order.  We did not parallelize the computations of $B_j$
and $B^{pref}_j$ because they turned out to represent only a tiny fraction of
the total preprocessing time.  

Finally, \name allocates the output
array $\tilde R_j$ and copies the input tuples from $R_j$ into their
respective partition $R_{j,s_1,\ldots,s_k}$ in $\tilde R_j$: notice
that this partition starts at position $B^{pref}_j[s_1,\ldots,s_k]$ in
$\tilde R_j$.  Copying is done in parallel, using a number of threads
equal to the number of available cores.  Each thread is responsible
for an exclusive subset of the $\Pi_j$ partitions: it scans the entire
array $A_j$ and, if the current tuple $(s_1, \ldots, s_k)$ belongs to
one of its partitions, then it physically copies the corresponding
$R_j$-tuple to the partition $R_{j,s_1,\ldots, s_k}$, otherwise it
ignores the tuple.  

There is no write contention between threads and
the work is uniformly distributed.  Since all threads need to read the
entire vector $A_j$, in order to minimize the total amount of work, we
restrict the number of threads to the number of physical cores. At
the end of this phase, each partition $R_{j,s_1,\ldots,s_k}$ is stored
in a subarray of $\tilde R_j$, starting at the position
$B^{pref}_j[s_1, \ldots, s_k]$.

\begin{example} \label{ex:partition} For a simple illustration,
  consider the Loomis-Whitney query,
  $Q=R_1(X, Y, Z), R_2(X, Y, U),$ $R_3(X, Z, U), R_4(Y, Z, U)$.  Assume
  the variable order is $X,Y,Z,U$ and
  $P = 2 \times 2 \times 1 \times 2 = 8$.

  In Fig.~\ref{fig:parapart}, we show a simple instance of the
  relation $R(X,Y,Z)$, the arrays $B, B^{pref}$ and the output
  $\tilde R$.  For illustration, we made some arbitrary assumptions
  about the hash functions, e.g. $h_Y(1)=1$, $h_Y(4)=h_Y(5)=h_Y(3)=1$,
  shown in the relation $A$.

  Each partition is color coded, and its tuples occur in a continuous
  subarray of $\tilde R$, e.g.  the grey partition with identifiers
  $0,1,0$ consists of $(0,4,1), (5,3,9)$, has size $B[0,1,0]=2$ and
  starts at position $B^{pref}[0,1,0]=3$ in $\tilde R$. The white
  Partition $1,0,0$ is empty: $B[1,0,0]=0$.
\end{example}

\begin{figure}[t]
  \centering
  \includegraphics[height=.33\textheight]{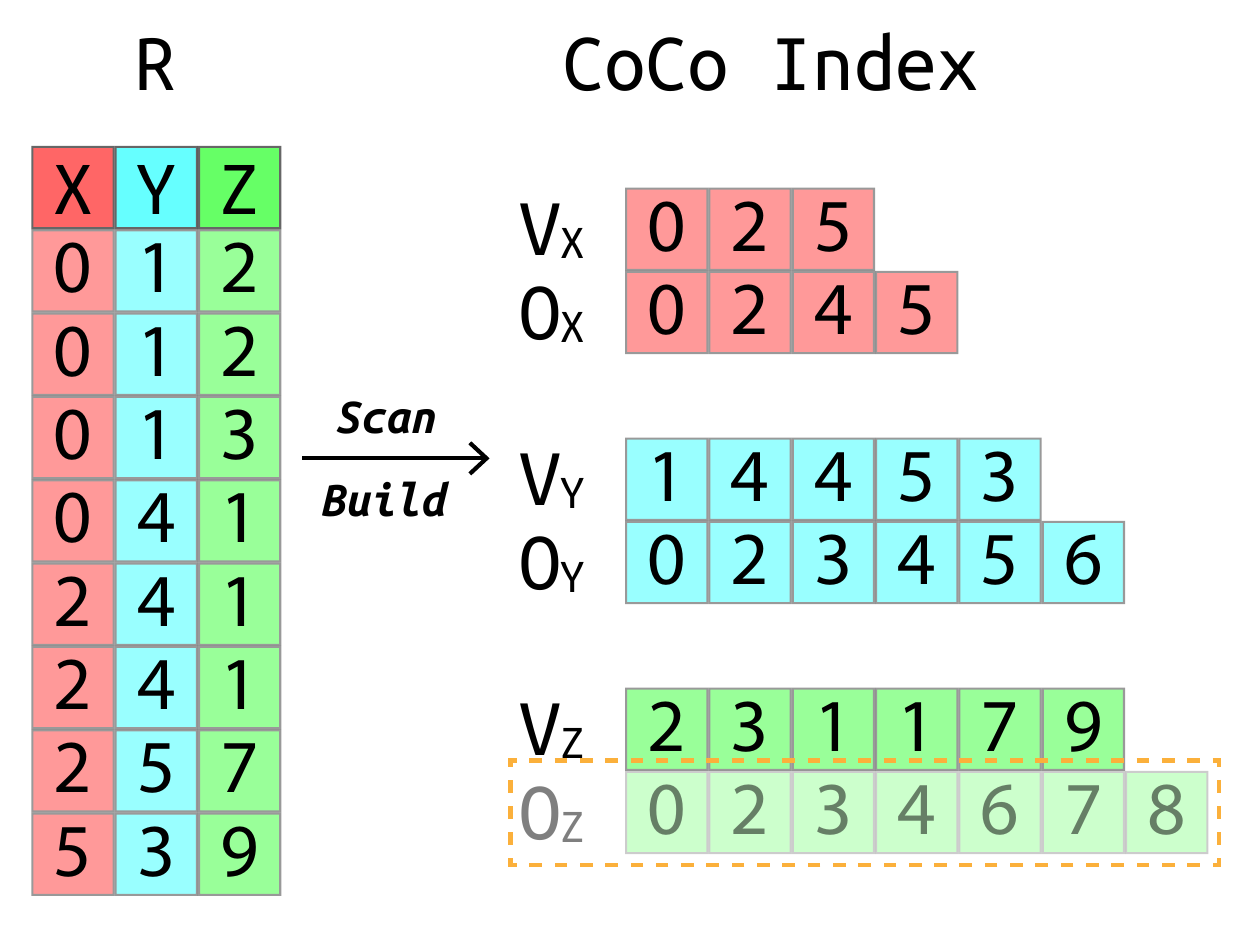}
  \caption{Construction of \indexlayout}
  \label{fig:cocoindex}
\end{figure}

Next, the preprocessor computes a trie index for each partition
$R_{j,s_1, \ldots, s_k}$ of each relation $R_j$. The index is a new
and simple data structure that we call \indexlayout (Compressed Column
layout), \revB{which combines ideas from both the trie in
  LFTJ~\cite{DBLP:conf/icdt/Veldhuizen14} and the memory formats used
  by sparse tensor
  compilers~\cite{DBLP:journals/pacmpl/ChouKA18}}. \indexlayout is
defined as follows. For a $k$-ary relation $R(X_1, \ldots, X_k)$,
\indexlayout consists of $k$ arrays $C_1, ..., C_{k}$.  Each array
$C_r$ contains the entire level $r$ of a sorted trie.  Its entries are
pairs $(x_i,p_i)$, where $x_i$ is a value of $R.X_i$ and $p_i$ is an
index to the beginning of a subarray of $C_{i+1}$.  The top vector
consists of all distinct values $x_1$ in $R.X_1$, sorted in ascending
order, and, for each pair $(x_1, p_1)$ in $C_1$, the index $p_1$
represents the beginning of a subarray in $C_2$ that contains all
distinct values $x_2$ in $R[x_1].X_2$.  And so on. The last vector
$C_{k}$ has pointers in the data array $R$. \name constructs a
separate \indexlayout index for each partition, $R_{j,s_1,...,s_k}$ of
the relation $R_j$.

To build \indexlayout, we sort the partitions $R_{j,s_1,\ldots,s_k}$
lexicographically, according to the variable order provided by the
optimizer: recall that we assumed, for simplicity, that the order is
$X_1, \ldots, X_K$. For example for the triangle query in
Fig.~\ref{fig:algo-comp}, if the order is $X,Y,Z$ then the relation
$T(Z,X)$ will be sorted by $X$ first then by $Z$.  Since each
partition $R_{j,s_1,\ldots,s_k}$ is a subarrays of a larger array
$\tilde R_j$, there are two possibilities to sort it.  One is to sort
each partition independently, in parallel threads; we do this when the
number of partitions $\Pi_j$ of $R_j$ is larger than the number of
available cores. If $\Pi_j$ is too small (e.g. when all shares $P_i$
of all variables in $R_j$ are $=1$), then we use a single parallel
sorting function and sort together the pair of arrays
$A_j,\tilde R_j$, thus forcing the tuples with the same partition to
stay together.  For this purpose we use \textbf{ips4o} (In-place
Parallel Super Scalar Samplesort)~\cite{axtmann2017,
  axtmann2020engineering}, which is a highly optimized parallel
sorting function for multicores.

Next, for each sorted partition $R_{j, s_1, ..., s_k}$, the
\indexlayout is constructed by compressing the same column values with
the same prefix determined by the variable order.  For each variable
$X_r$, the array $C_r$ consists of two separate arrays $V_r, O_r$,
where $V_r$ for the values of $X_r$ and $O_r$ for offsets into
$C_{r+1}$.  

We first scan over all rows in the partition by computing
the number $l_r$ of unique values with prefix $(\ldots, s_r)$ for any
attribute $X_r$. Then, we allocate the space with $l_r$ size for $V_r$
and $l_{r}+1$ size for $O_r$ and scan again to populate them from
bottom to above. During the second pass, we bookkeep a cursor about
current offset $o_i$ in the compressed array for each attribute and
sequentially compare the adjacent two rows $(s_1, \ldots, s_{k})$ and
$(t_1, \ldots, t_{k})$ and find the index $e$ of the first differing
element $s_e \neq t_e$ but
$(s_1, \ldots, s_{e-1}) = (t_1, \ldots, t_{e-1})$. Then, we create a
new entry in $V_e$ with the value $t_e$ and the offset $o_e$ in $O_e$,
and increment $o_e$. Moreover, since the prefix is changed, we need to
create entries with value $t_d$ and offset $o_d$ for any $V_d, O_d$
with any $d > e$ and increment $o_d$ as well.  Besides, if the
relation is unique, we will ignore the offset vector of last attribute
$O_k$ and its construction.

\begin{example}
  In Fig.~\ref{fig:cocoindex}, we show the \indexlayout index for the relation $R(X, Y, Z)$.\footnote{It ought to be an index on partition, but for simplicity, just on the whole relation.} 
  The first vector $C_1 = C_X$ contains all distinct values $V_X$ of $X$ and the offsets $O_X$ in the $C_Y$ (starts with $0$). The second vector $C_2 = C_Y$ depends on the prefix of $X$ and contains all values $V_Y$ of $Y$ with different $X$, thus we can see two $4$ in the $V_Y$ since they have different $X$ values. 
  The last vector $V_3 = V_Z$ is just similar cases, but the last level offsets $O_3 = O_Z$ will be removed if the relation $R$ itself is distinct.
\end{example}

\section{Executor}

\label{sec:join}

\begin{figure}[t]
  \centering
  \includegraphics[height=.33\textheight]{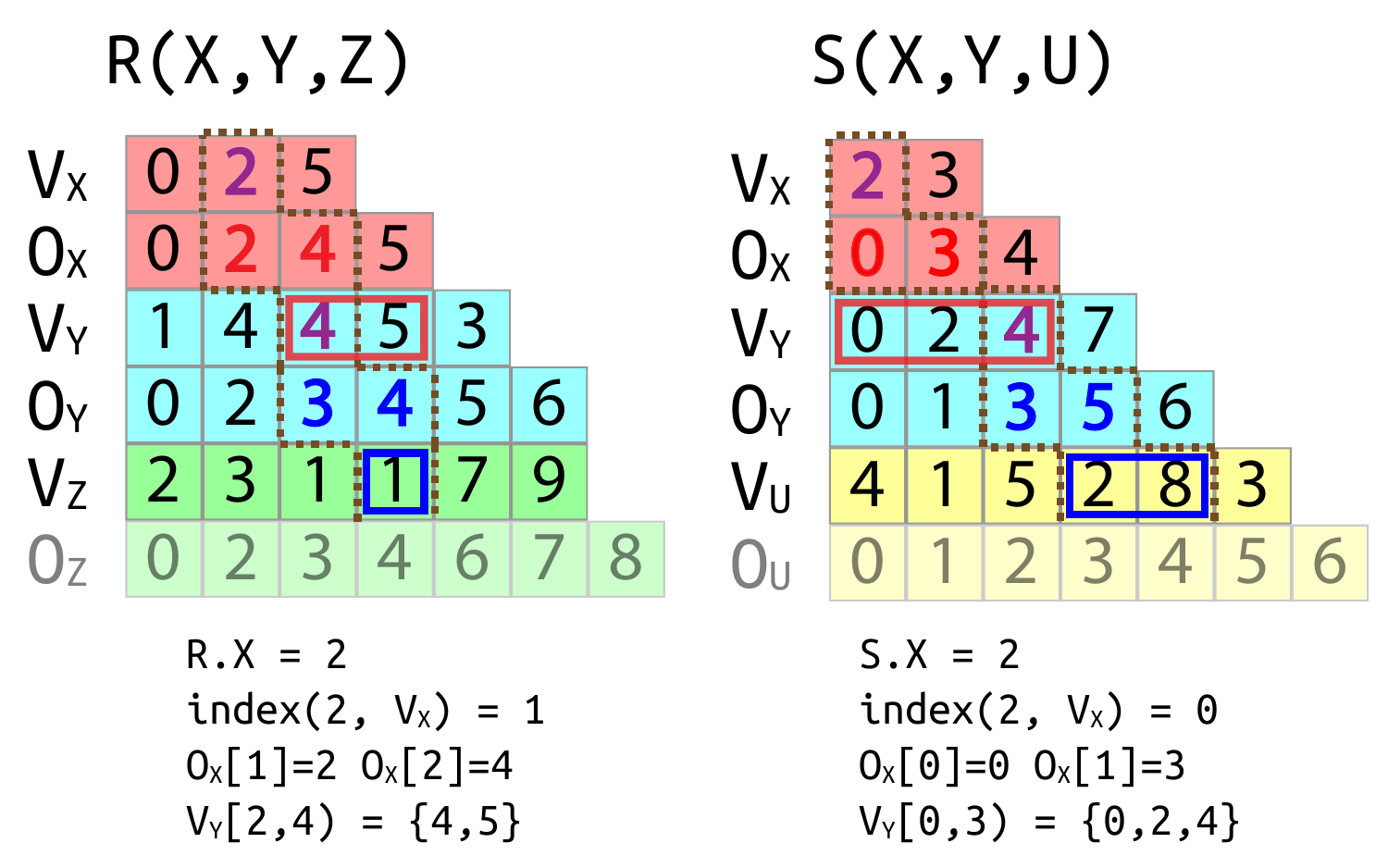}
  \caption{Join Execution on \indexlayout}
  \label{fig:cocojoin}
\end{figure}

After preprocessing, the Executor computes Generic Join, in parallel.
Recall that the query $Q$ in Eq.~\eqref{eq:full:cq} has $n$ variables
$X_1, \ldots, X_n$.  The optimizer has computed the shares
$P_1, \ldots, P_n$ for each variable, and has chosen a variable order,
which we assume w.l.o.g. is $X_1, \ldots, X_n$, while the preprocessor
has partitioned each input relation $R_j$ and moved it to a new array
$\tilde R_j$.

The executor assigns each partition to a thread with identifier
$(s_1, s_2, \ldots, s_n)$.  Each thread executes Generic Join
independently, with the only difference that, for each relation
$R_j(X_{i_1}, \ldots, X_{i_k})$, it only accesses the subarray of
$\tilde R_j$ corresponding to the partition
$R_{j,s_{i_1}, \ldots, s_{i_k}}$, which starts at offset
$C_j[s_{i_1}, \ldots, s_{i_k}]$ in $\tilde R_j$.  Otherwise, the
structure of the executor is the same as that of the standard generic
join: a sequence of nested loops, each corresponding to a variable
$X_i$.  The main work is done by the intersection
$R_{j_1}.X_i \cap R_{j_2}.X_i\cap \cdots$, which we describe in detail
next.  Notice that the total number of threads is the total number of
output partitions, which is $P_1 \times P_2 \times \cdots \times P_n= P$.  Also,
there is no read-write conflict between threads: their only
interaction is that they read from the shared memory.  Besides, due to
non-overlapping join results, each thread can store its outputs in its
local memory; at the end of the computation, these results can be
concatenated if needed.

Listing~\ref{lst:newpara} in Fig.~\ref{fig:algo-comp} shows the execution performed by the
thread with the identifier $(i,j,k)$; this thread only needs to access the
partitions $R_{ij}, S_{jk}$, and $T_{ik}$, and join them.

Next, we present the details of the intersection, which is the main
work done in WCOJ. By using \indexlayout, the intersection needs to compute common values in a set of sorted and contiguous
arrays.  For this purpose, \name uses ideas from multi-way merge.

Given a sorted array $V$ and a value $x$, the method
$\text{search}(V, x)$ (described below) that finds the position of the
first element in $V$ that is $\geq x$.  To compute the intersection of
multiple arrays $V_1 \cap V_2 \cap \cdots$ we use a cursor in each
array, and maintain the minimum $\text{min}$ and maximum $\text{max}$
values among all cursors.  If $\text{max} \neq \text{min}$, we just
advance the cursor of $\text{min}$ to
$\text{search}(V_j, \text{max})$; otherwise, we have found a common 
value $\text{min}=\text{max}$.

For the first attribute $X_1$ in variable order, the intersection
$R_{j_1}.X_1 \cap R_{j_2}.X_1 \cap \cdots$ is done by the above
procedure on the first levels $V_1$ in \indexlayout of
$R_{j_1}, R_{j_2}, \ldots$.  Once we encounter a value $X_1 = x$
present in all relation, for each relation $R_{j_k}$ we need to
restrict the range of next trie level.  Assuming that $x$ occurs on
position $\mathrm{index}$ in $V_1$, then the next level will be
restricted to the subarray
$V_2[O_1[\mathrm{index}]: O_1[\mathrm{index}+1]-1]$.  In general, each
intersection will be done by intersecting several arrays, which can be
either a full \indexlayout array $V_r$, or a subarray thereof.

\begin{example}
  In Fig.~\ref{fig:cocojoin}, The area surrounded by the brown line shows a state during the join between $R(X, Y, Z)$ and $S(X, Y, U)$ on the \indexlayout. Now, we already find that $R.X = S.X = 2$ is an intersection on the attribute $X$. To find the intersecting range of $Y$ in $R$, we need compute the $\mathrm{index}(R.X, V_X) = 1$, and obtain the range $[O_X[1], O_X[2]) = [2, 4)$ in the $V_Y$, which is the subarray $\{4, 5\}$. The similar case happened on $S.Y$ and obtained the subarray $\{0, 2, 4\}$. Then we do the intersection on two subarrays and find the intersected value $Y = 4$.
\end{example}

To further improve the intersections on sorted arrays efficiently, 
we propose three $search$ ways, which are Exponential, Quadratic, and Linear, 
and use them based on the number of attributes involved in the intersection. 
Exponential Search~\cite{DBLP:conf/birthday/Baeza-YatesS10}, also known as galloping search, 
is suitable when dealing with skewed and large datasets. 
It begins by jumping exponentially (i.e., $2^n$) through the data until it overshoots the target value, 
after which it switches to a binary search within the identified range. 
Quadratic Search is somehow used for medium-sized datasets 
and involves first a series of steps that increase by fixed size 
which is set to the multiple of cache line size, 
and then scan within identified range or recursively apply Quadratic search (but unrolling in implementation). 
Finally, Linear Search is employed for small datasets, 
as it is the most efficient search method for a small number of elements which is fitted the cache size.
By selecting the appropriate search method based on the size of the intersection, 
we ensure that the join operation is optimized for each scenario, maximizing efficiency and performance.

\section{Optimization}\label{sec:opt}

The optimizer in \name has three goals: it chooses shares
$P_1, \ldots, P_n$ for the variables $X_1, \ldots, X_n$, it chooses a
variable order $X_{\sigma(1)}, \ldots, X_{\sigma(n)}$ to be used by
the query executor, and, finally, it performs a rewriting of the code
of Generic Join in order to remove redundant computations inherent in
this algorithm.  All decisions are done jointly, and are informed by a
cost model described in Sec.~\ref{subsec:cost:model}.
Although it happens last, we start our presentation of the optimizer
by describing query rewriting.

\subsection{Query Rewriting}
\label{subsec:rewrite}

Recall that Generic Join consists of $n$ nested loops, one loop for
each variable $X_i$.  The actual work in GJ consists of computing the
intersection of all columns of the variables $X_i$.  For complex
queries, some of these intersections computed in the inner loops are
independent of outer loop and, thus, are repeated unnecessarily for
each iteration of the outer loop.  In this case our optimizer rewrites
the query such as to compute the intersection early, and stores this
as a temporary result.  We illustrate query rewriting with an example;
the general case follows immediately.

\begin{figure}[t]
  
{
  \begin{minipage}{0.495\textwidth}
\begin{lstlisting}[style=BashInputStyle,
  label=lst:beforeopt]
Q(X,Y,Z,U) = R1(X,Y), R2(X,Z), R3(X,U),
              $\,$R4(Y,Z), R5(Y,U), R6(Z,U).

// $\textbf{BEFORE REWRITING}$
For x $\in$ R1.X $\cap$ R2.X $\cap$ R3.X
  For y $\in$ R1[x].Y $\cap$ !\colorbox{green}{R4.Y $\cap$ R5.Y}!
    For z $\in$ !\colorbox{yellow}{R2[x].Z}! $\cap$ R4[y].Z $\cap$ !\colorbox{yellow}{R6.Z}!
      For u $\in$ !\colorbox{pink}{R3[x].U $\cap$ R5[y].U}! $\cap$ R6[z].U
        Q += (x, y, z, u)
\end{lstlisting}
  \end{minipage}
  
  \begin{minipage}{0.495\textwidth}
\begin{lstlisting}[style=BashInputStyle,
  label=lst:afteropt]  
// $\textbf{AFTER REWRITING}$ 
tmp_Y := !\colorbox{green}{R4.Y $\cap$ R5.Y}! // Lift Up Y
For x $\in$ R1.X $\cap$ R2.X $\cap$ R3.X
  tmp_Z := !\colorbox{yellow}{R2[x].Z $\cap$ R6.Z}! // Lift Up Z
  For y $\in$ R1[x].Y $\cap$ !\colorbox{green}{tmp\_Y}!
  tmp_U := !\colorbox{pink}{R3[x].U $\cap$ R5[y].U}! // Lift Up U
    For z $\in$ R4[y].Z $\cap$ !\colorbox{yellow}{tmp\_Z}!
      For u $\in$ R6[z].U $\cap$ !\colorbox{pink}{tmp\_U}!
        Q += (x, y, z, u)
\end{lstlisting}
  \end{minipage}
}
\caption{Example of Query Rewriting}
\label{fig:rewrtie}
\end{figure}

\begin{example}[Duplicated Intersections] \label{ex:duplicate:intersections}
Consider the following query which computes a 4-clique $Q$ on the variables $(X,Y,Z,U)$. Generic Join for $Q$ is shown in Fig~\ref{fig:rewrtie} (left), with three pieces of redundant
work highlighted. They do not depend on the current loop variable, and
therefore are computed repeatedly:

\revB{Here $R4[y].Z$ represents the column  $Z$ of the subset of $R4$ where $Y=y$.}

For example, R4.Y $\cap$ R5.Y is computed repeatedly, for every value
$x$, although this expression does not depend on $x$.

\end{example}

\revB{Given the variable order, our optimizer identifies 
  intersections  that can be  cached, then}

proceeds to "lift" these intersections, 
treating them as independent computational entities. 
These independent intersections are then computed only once before the iteration over the current loop variable and cached within auxiliary data structures. 
This process not only reduces the need for repeated calculations 
but also ensures that these pre-computed intersections 
can be efficiently accessed and reused across different parts of the join operation.

\begin{example}  \label{ex:duplicate:intersections:2}
  The optimizer rewrites the code above as follows:

\revB{ Instead of computing the intersection
  $R1[x].Y \cap R4.Y \cap R5.Y$ for each $x$, we compute
  $R4.Y \cap R5.Y$ before starting the iteration on $x$ and cache the
  result in a temporary sorted array \texttt{tmp\_Y}, as shown in Fig~\ref{fig:rewrtie} (right).  Then, during
  the iteration over $x$, we intersect $R1[x].Y$ with \texttt{tmp\_Y}.
  Importantly, the temporary array \texttt{tmp\_Y} also stores offsets
  corresponding to the value $y$ in the \indexlayout of $R4$ and $R5$.
  
  While this optimization does not improve the asymptotic runtime of
  WCOJ, it can improve the actual runtime.  To see this, consider how
  WCOJ computes the intersection $R1[x].Y \cap R4.Y \cap R5.Y$.
  Denoting the three sorted arrays $R1[x].Y$, $R4.Y$, $R5.Y$ by
  $A,B,C$ respectively, the simplified pseudocode is shown in
  Fig.~\ref{fig:multiwayjoin}: notice that \name uses exponential search
  instead of the linear search, see Sec.~\ref{sec:join}.  By
  precomputing the intersection of $B \cap C$, we replace a 3-way
  merge with two 2-way joins, which avoids repeating the same
  iterations over $B, C$ for every value of $x$, and also have a
  simpler code and better cache locality.  }

\begin{figure}[t]
  
{

  \begin{minipage}{0.495\textwidth}
    \begin{lstlisting}[style=BashInputStyle, label=lst:multijoin]
i = j = k = 0;
while not done:
  if A[i] = B[j] = C[k]: 
    output(A[i]); 
    i++, j++, k++;
  while (A[i] < min(B[j], C[k])):
    i++;
  while (B[j] < min(A[i], C[k])):
    j++;
  while (C[k] < min(A[i], B[j])):
    k++;
\end{lstlisting}
    \end{minipage}
    
    \begin{minipage}{0.495\textwidth}
\begin{lstlisting}[style=BashInputStyle, label=lst:multijoin]
j = k = p = 0;
while not done:
  if B[j] = C[k]: 
    Tmp[p++] = B[j++]; j++, k++;
  while (B[j] < C[k]): j++;
  while (C[k] < B[j]): k++;
 . . . .
i = p = 0
while not done:
  if A[i] = Tmp[p]: 
    output(A[i]); i++, p++;
  while (A[i] < Tmp[p]): i++;
  while (Tmp[[p] < Ai]): p++;
\end{lstlisting}
    \end{minipage}}

  \caption{\revB{Simplified Pseudocode for computing
      $A \cap B \cap C$, and same code that caches $B \cap C$ before
      computing the intersection.  \name uses exponential search
      (Sec.~\ref{sec:join}) instead of the linear search shown here.
      Notice that these two code fragments are equivalent only if
      $A,B,C$ are sorted arrays.}}
    \label{fig:multiwayjoin}
\end{figure}

\revB{ We note that this query rewriting is a logical optimization,
  and is not an optimization that can be done by a compiler, e.g. by
  using LLVM Loop Invariant Code Motion (-licm) pass. A simple reason
  is that there is no piece of code in Fig.~\ref{fig:multiwayjoin}
  (left) that represents $B \cap C$, which the compiler could attempt
  to move.  A deeper reason is that the optimization in
  Fig.~\ref{fig:multiwayjoin} is sound only if the arrays $A,B,C$ are
  sorted, which is an invariant that compilers usually cannot infer.

}
\end{example}

\name stores each temporary intersection locally in each thread.
There are no read-write or write-write conflicts between threads
during the parallel execution.  Each thread only operates on its
locally cached data, and there is no need for a lock-based
synchronization.  Therefore this optimization reduces duplicated work,
without creating additional bottlenecks.

\revB{ Next, we discuss the complexity of the optimized algorithm.  A
  \emph{fractional edge cover} of the full conjunctive query in
  Eq.~\eqref{eq:full:cq} is a tuple of non-negative numbers
  $\bm w = (w_1, \ldots, w_m)$ such that ``every variable $X$ is
  covered'', meaning $\sum_{j: X \in \text{Vars}(R_j)} w_j \geq 1$.
  It is known that, for any fractional edge cover $\bm w$, Generic
  Join runs in time $\tilde O(\prod_j |R_j|^{w_j})$.  Consider now an
  optimized algorithm $P$, where some intersections have been cached
  early, and let $Q_P$ the query obtained from $Q$ as follows: $Q_P$
  has the same atoms $R_j$ as $Q$, and each variable $X$ occurs only
  in those atoms $R_j$ that are used in the first intersection of the
  $X$ domains. We prove in the full version of the paper:\footnote{By
    a simple adaptation of the optimality proof of generic
    join~\cite{DBLP:journals/sigmod/NgoRR13}.}

\begin{theorem}
  For any fractional edge cover $\bm w$ of $Q_P$, algorithm $P$ runs in time
  $\tilde O(\prod_j |R_j|^{w_j})$.
\end{theorem}

To illustrate, consider the query $Q$ in
Example~\ref{ex:duplicate:intersections}.  If $R_2, R_5$ are the
smallest of the 6 relations, then standard Generic Join runs in time
$\tilde O(|R_2|\cdot |R_5|)$, because of the fractional edge cover
$\bm w = (0, 1, 0, 0, 1, 0)$.  The query $Q_p$ associated to the
program $P$ in Example~\ref{ex:duplicate:intersections:2} is
$Q_P = R_1(X) \wedge R_2(X,Z) \wedge R_3(X,U) \wedge R_4(Y) \wedge
R_5(Y,U) \wedge R_6(Z)$ ($Y$ only occurs in $R_4, R_5$ because of
$\text{tmp}_Y := R_4.Y\cap R_5.Y$).  Since $\bm w$ is also a
fractional edge cover of $Q_P$, algorithm $P$ runs in optimal time
$\tilde O(|R_2|\cdot |R_5|)$; in particular, $P$ is optimal when
$|R_1|=\cdots = |R_6|=N$.  However, if $R_1, R_6$ are strictly smaller
than all other relations, then $P$ is no longer optimal.

If theoretical optimality is required, it is possible to check for
each candidate rewriting $P$ if its complexity is the same as that of
standard generic join.  In \name we use a cost model instead, as we
describe next.}

\subsection{Cost Model}

\label{subsec:cost:model}

We describe here our model for estimating the cost of Generic Join on the
query $Q$ in~\eqref{eq:full:cq}, assuming a variable order
$\bm X_\sigma$ given by a candidate permutation $\sigma$, i.e.
$X_{\sigma(1)}, X_{\sigma(2)}, \ldots, X_{\sigma(n)}$; refer to the
notations in Table~\ref{tab:freq}.  

Consider some level $i=1,n$.  The outer iterations have bound their
variables to the prefix
$x_{\sigma(1:i-1)}\defeq
(x_{\sigma(1)},x_{\sigma(2)},\ldots,x_{\sigma(i-1)})$.  Let
$\SetS[X_{\sigma(i)}]=\set{R_{j_1}, R_{j_2}, \ldots}$ (or just $\SetS$
when the variable $X_{\sigma(i)}$ is clear from the context) be the
set of relation names that contain the variable $X_{\sigma(i)}$, and
$|\SetS|$ be its size. Then the $i$'th loop needs to compute
$R_{j_1}[x_{\sigma(1:i-1)}].X_{\sigma(i)} \cap
R_{j_2}[x_{\sigma(1:i-1)}].X_{\sigma(i)}\cap \cdots$.  We denote by
$\SetN[x_{\sigma(1:i-1)}\oplus X_{\sigma(i)}]$ the set of
cardinalities of the relations in $\SetS$, i.e.
$\set{|R_{j_1}[x_{\sigma(1:i-1)}]|, |R_{j_2}[x_{\sigma(1:i-1)}]|,
  \ldots}$, and by $\min \SetN$, $\max \SetN$ the smallest/largest
cardinality.  
For example, consider the triangle query in
Fig.~\ref{fig:algo-comp}, where the variable order is $X,Y,Z$.  In the
second loop ($\texttt{for } y \texttt{ in} \ldots$), we have
$\SetS = \SetS[x\oplus Y] = \set{R[x], S}$, $|\SetS|=2$, and
$\SetN[x\oplus Y]$ is $\set{|R[x]|, |S|}$. If $|R[x]| < |S|$, then $\min \SetN = |R[x]|$ and $\max \SetN = |S|$.

Next, we describe the cost of computing the intersection at level $i$,
which we denote by
$\Restrict{\Cost}{\Concat{\ctxpp{1:i-1}}{\VTXPP{i}}}$.  Assume that
the intersection is computed using galloping search: iterate over the
smallest relation $R_{j_{\min}}$ with the cardinality $\min \SetN[x_{\sigma(1:i-1)}\oplus X_{\sigma(i)}]$,  and probe using exponential search
in each of the remaining relations $R_j$.  If the values in
$R_{j_{\min}}$ are uniformly distributed in $R_j$, then the cost is
$\log\frac{|R_j|}{|R_{j_{\min}}|}$, which leads to:

\begin{equation*}
\Restrict{\Cost}{\Concat{\ctxpp{1:i-1}}{\VTXPP{i}}} = |\Restrict{\SetS}{\VTXPP{i}}|\cdot \min \Restrict{\SetN}{\Concat{\ctxpp{1:i-1}}{\VTXPP{i}}} \cdot \log_2\Big(1 + \frac{\max \Restrict{\SetN}{\Concat{\ctxpp{1:i-1}}{\VTXPP{i}}}}{\min \Restrict{\SetN}{\Concat{\ctxpp{1:i-1}}{\VTXPP{i}}}}\Big)
\end{equation*}

So far the cost is for a single binding $\bm x_{\sigma(1:i-1)}$ of the
prefix $\bm X_{\sigma(1:i-1)}$ of the variables.  To add up their
cost, we need a notation.  Consider our definition of the conjunctive
query in Eq.~\eqref{eq:full:cq}: the head variables are
$X_1, \ldots, X_n$.  We denote by $Q(\bm X_{\sigma(1:i-1)})$ the query
where the head variables are restricted to $\bm X_{\sigma(1:i-1)}$.
The bindings  $\bm x_{\sigma(1:i-1)}$ will iterate over all outputs of
this query, hence the total cost at the level $i$ is:

\begin{equation*}
  \Restrict{\Cost}{\VTXPP{1:i}} = \qquad \smashoperator[lr]{\sum_{\ctxpp{1:i-1} \in Q(\bm X_{\sigma(1:i-1)})}} \qquad \Restrict{\Cost}{\Concat{\ctxpp{1:i-1}}{\VTXPP{i}}} 
\end{equation*}

Ultimately, we define the total cost $\TotalCost[\VTXP]$ of our join algorithm by summing up the intersection costs for all variables in the variable order $\VTXP$. 
\begin{equation*}
  \TotalCost[\VTXP] = \TotalCost[\VTXPP{1:n}] = \smashoperator[lr]{\sum_{i \in [n]}} \Restrict{\Cost}{\VTXPP{1:i}} 
\end{equation*}

This cost cannot be computed exactly at optimization time, instead we
compute an upper bound.  For that purpose we use a pessimistic
cardinality
estimator~\cite{DBLP:conf/sigmod/CaiBS19,DBLP:journals/pacmmod/KhamisNOS24,10.1145/3651597}
to upper bound the output size of the
queries $Q(\bm X_{\sigma(1:i-1)})$, and use the following inequality,
which we prove in the full version of the paper:

\begin{lemma}\label{lem:cost-ub}
The following inequality holds, 
where the summation is over $\bm x_{\sigma(1:i-1)} \in Q(\bm
X_{\sigma(1:i-1)})$, and we drop the argument
$\Restrict{\cdot}{\Concat{\ctxpp{1:i-1}}{\VTXPP{i}}}$ from $\SetN$:
  \begin{equation}
    \Restrict{\Cost}{\VTXPP{1:i}} \leq |\SetS(X_{\sigma(i)}| \cdot (\sum \min \SetN) \cdot \log_2\Big(1 + \frac{\sum \max \SetN}{\sum \min \SetN}\Big)\label{eq:final:cost}
  \end{equation}
\end{lemma}

\begin{example}
  Referring again to the 2nd loop of the triangle query,
  \revB{$R[x].Y\cap S.Y$}, we have $\SetN = \set{|R[x]|, |S|}$.  We
  compute Eq.~\eqref{eq:final:cost} as follows.  \revB{First,
    $\sum \min \SetN$ is $\sum_x \min(|R[x].Y|,|S.Y|)$, which is the
    same as the output size of the query $Q(X,Y)=R(X,Y),S(Y),T(X)$
    (where $S(Y),T(X)$ represent $\Pi_Y(S)$ and $\Pi_X(T)$
    respectively). We upper bound it using a pessimistic
    cardinality estimator: \name uses~\cite{10.1145/3651597} for this
    purpose.  For the expression inside the logarithm 
    $\dfrac{\sum_x\max(|R[x].Y|,|S.Y|)}{\sum_x
      \min(|R[x].Y|,|S.Y|)} = \dfrac{\avg_x\max(|R[x].Y|,|S.Y|)}{\avg_x
      \min(|R[x].Y|,|S.Y|)}$.  Since this does not have a
    closed form, we estimate it as
    $\dfrac{\max(\avg_x(|R[x].Y|),\avg_x(|S.Y|))}{\min(\avg_x(|R[x].Y|),\avg_x(|S.Y|))}$.
    Next, we compute $\avg_x(|R[x].Y|) = |R|/|R.X|$, and similarly for
    the other terms. 
    The 1st and the 3rd loops of the triangle query are handled similarly.
    }

\end{example}

If an intersection at level $i$ is rewritten to be computed at an
earlier level $i_0<i$ (Sec.~\ref{subsec:rewrite}), then those
relations will be included in the set $\SetS(X_{\sigma(i_0)})$, while
the temporary relation will be added to $\SetS(X_{\sigma(i)})$.

\subsection{Plan Decision} \label{subsec:plan}

We describe now how the optimizer chooses the optimal variable order,
$\bm X_\sigma$ and shares $P_1 \times P_2 \times \cdots \times P_n = P$, 
where $P_i$ represents the share of the variable $X_{\sigma(i)}$.

The HyperCube partition leads implicitly to some replicated work,
which we capture by the following expressions:

\begin{equation}
  \Restrict{\TotalCost}{\VTXP, \PartSharePerm} = \smashoperator[lr]{\sum_{i \in [n]}} \Big((\prod_{j > i} P_{\sigma(j)}) \Restrict{\Cost}{\VTXPP{1:i}} \Big)\label{eq:total:cost}
\end{equation}

For example, in the triangle query, for a fixed $x$, every value
$y \in R[x].Y\cap S.Y$ is discovered redundantly by all threads with
partition identifiers the form $(i,j,*)$ where $i=h_X(x)$ and
$j=h_Y(y)$: this accounts for the product $\prod_{j>i}$ above.

A naive way to compute the optimal variable order $\sigma$ and shares
$P_i$ is to try all combinations and return the cheapest
cost~\eqref{eq:total:cost}.  There are $n!$ variable orders and
${10+n-1\choose n-1}$ total possible share allocations (since we use by
default $P=1024=2^{10}$ threads).  A brute force search only works for
small values of $n$.  Instead, we apply some pruning heuristics, as follows.

We prune partition shares that increase the cost
$\Restrict{\TotalCost}{\VTXP, \PartSharePerm} > 2 \cdot
\Restrict{\TotalCost}{\VTXP}$.  We will also prune small partitions.
Recall that if we throw randomly $B$ balls into $P$ bins and
$B = O(N)$, then the expected size of the largest bin is not
$O(B/P)=O(1)$, but it is $O(\log P)$, which means that the data is
non-uniformly distributed.  To expect uniform distribution, $O(B/P)$,
we need $B > 3 P \log P$ balls~\cite{DBLP:journals/sigmod/KoutrisS16}.
Therefore, we prune partitions containing some share $P_i$ where
$|\dom(X_i)| < 3 \cdot P_i \cdot \log P_i$

Finally, we assign an \emph{eveness} score $E$ to a set of shares,
which favors evenly distributed, with a bias towards have more shares
for last variables $X_{\sigma(n)}, X_{\sigma(n-1)}, \ldots$ than for
the first variables.  $E$ is defined as:

\begin{equation}
  E(\PartSharePerm) = \sum_{i \in [n]} P_{\sigma(i)} \cdot w(i)
\end{equation}
where $w(i) = \max\{1 - \frac{i}{100}, \frac{3}{4}\}$ is a smooth weight function on $i$ to distinguish the importance of the variable. Finally, we choose the partition share with the minimum evenness among all feasible partitions, to balance the task computations and avoid data skewness.
\section{Experiments} \label{sec:exp}

We implemented \name as a standalone C++ program. It reads the query
and the data path from a JSON file, and loads data from a binary file,
which was previously converted from CSV format. After running the
query, the program can either output all result tuples, or output only
the count.  We compared \name against some state-of-the-art systems
that support cyclic join queries, and conducted some scalability and
ablation studies. We ask four research questions:

\begin{enumerate}
	\item How does the absolute performance of \name compare to that of related systems on both real and synthetic datasets?
	\item How does \name scale up with an progressively increased number of cores?
	\item How important is it to optimize the partition shares and the variable order?
	\item How effective is the query rewriting method and the \indexlayout data structure?
\end{enumerate}

\subsection{Setup}

{
\begin{table}[t]
	\caption{Dataset Characteristics}
	\label{tab:data}
	\begin{tabular}{ccll}
		\toprule
		Name & \# Node & \# Edge & Feature \\
		\midrule
		WGPB~\cite{wgpb-dataset, DBLP:conf/semweb/HoganRRS19} & 54.0M & 81.4M & sparse, skew   \\
		Orkut~\cite{DBLP:conf/imc/MisloveMGDB07} & $3.07M$ & $117M$ & partial dense, uniform  \\
		GPlus~\cite{DBLP:conf/nips/McAuleyL12} & 107K & 13.6M & dense, skew \\
		USPatent~\cite{DBLP:conf/kdd/LeskovecKF05} & 3.77M & 16.5M & sparse, uniform \\
		Skitter~\cite{DBLP:conf/kdd/LeskovecKF05} & 1.69M & 11.1M & sparse, partial skew \\
		Topcats~\cite{DBLP:conf/kdd/YinBLG17} & 1.79M & 28.5M & partial dense, skew \\
		\bottomrule
	\end{tabular}
\end{table}
}

\noindent\textbf{Datasets:} We provide the information for our graph datasets in Table~\ref{tab:data}. These datasets include a variety of types, sparsity, and skewness, which cover the common characteristics of real-world datasets.

We transform those datasets into CSV format and either load them directly or use the provided loader to fetch them to the main memory. 

In addition, we also use two synthetic datasets for tensor kernels, with different sparsity. We generate the synthetic datasets with the same number of nodes and hyperedges where the number of attributes is set to 3. The first dataset (named ST-Dense) is generated uniformly densely distributed. The dataset consists of 7 relations, each with three attributes. The first four relations follow the form\footnote{$[N]$ represents all positive integers $\leq N$, $\times$ denotes the Cartesian product.} $[512] \times [512] \times [512]$. The last three relations have attribute domains $(\{0, 1\} | \{0, 2\} | \{1, 2\}) \times [4096] \times [4096]$. The second dataset (named ST-Sparse) is similar but just distributed sparsely. The first four relation randonly selected $512^3$ tuples from $[5120] \times [5120] \times [5120]$, the sparsity is $0.001$. The last three relations are selected $4096^2$ tuples from $[40960] \times [40960]$ while keeping the first attribute, thus the sparsity is $0.01$.

\addvspace{\smallskipamount}
\noindent\textbf{Baselines:} We compare our method with five state-of-the-art baselines: Umbra~\cite{DBLP:journals/pvldb/FreitagBSKN20}, Diamond~\cite{DBLP:journals/pvldb/BirlerKN24}, DuckDb~\cite{DBLP:conf/cidr/RaasveldtM20}, Soufflé~\cite{DBLP:conf/lopstr/ArchHZSS22}, and Kùzu~\cite{DBLP:conf/cidr/JinFCLS23}. 

Umbra (2022-11-03) is a high-performance database management system designed for modern hardware architectures. It nicely supports the worst-case optimal join (multiway join) algorithms in parallel through morsel-driven execution.
\revB{Diamond is a variant of Umbra, which is optimized to solve the
  diamond\footnote{Intermediate results are larger than the
  inputs and the output.} problem. It splits join operators into Lookup and
  Expand suboperators, and allows them to be freely reordered. It
  also uses a novel ternary operator  Expand3 that implements the
  triangle query.}
Kùzu (v0.6.0) is a graph database engine that excels at handling
complex queries involving joins across graph-structured data. Its most
recent version it supports  worst-case optimal join through ``join hints''~\cite{kuzujoinhint}.
Soufflé (v2.4.1) is a logic-based database engine that compiles queries into optimized C++ code. It supports efficient execution of Datalog queries, particularly for complex join operations, but does not implement worst-case optimal join.
DuckDB (v1.1.1) is an in-process analytical database management system optimized for fast OLAP workloads. It is heavily optimized for complex join queries, but does not support worst-case optimal join.

Umbra is provided in binary format and was obtained directly from the author, while the other baselines are open-source and were downloaded directly from their official GitHub repositories.

For Umbra, we enabled paged storage optimization, and forced it to use worst-case optimal join, which improved its performance. For Soufflé, we enabled brie representation for dense input relation, which also improved its performance. For Kùzu, we provided join order hints in order to facilitate the worst-case optimal join plan. All these non-default settings were chosen to improve the systems' overall performance. 

{
\begin{table*}[t]
	\centering
	\caption{Table of Queries}
	\label{tab:query}
	\begin{tabular}{ll}  
	\toprule
	\textbf{Name} & \textbf{Queries} \\ 
	\midrule
	Q1 (Triangle) & \texttt{Q(X,Y,Z) := R(X,Y), S(Y,Z), T(X,Z).}
	 \\ 
	\hline
	Q2 (4-Loop) & \texttt{Q(X,Y,Z,U) := R1(X,Y), R2(X,Z), R3(Y,U), R4(Z,U).} \\ 
	\hline
	Q4 (4-Diamond) & \texttt{Q(X,Y,Z,U) := R1(X,Y), R2(X,Z), R3(Y,U), R4(Z,U), R5(Y,Z).} \\ 
	\hline
	Q6 (4-Clique) & \texttt{Q(X,Y,Z,U) := R1(X,Y), R2(X,Z), R3(Y,U), R4(Z,U), R5(Y,Z), R6(X,U).} \\ 
	\hline
	Q8 (2-Triangle) & \texttt{Q(X,Y,Z,U,V) := R1(X,Y), R2(X,Z), R3(Y,Z), R4(Z,U), R5(Z,V), R6(U,V).}\\ 
	\hline
	\hline
	LW(Loomis-Whitney) \!\!\!& \texttt{Q(X,Y,Z,U) := R1(X,Y,Z), R2(X,Y,U), R3(X,Z,U), R4(Y,Z,U).}\\
	\hline
	CT (Clover-Triangle) & \texttt{Q(U,X,Y,Z) := R5(U,X,Y), R6(U,X,Z), R7(U,Y,Z).} \\
	\bottomrule
	\end{tabular}
\end{table*}
}

\addvspace{\smallskipamount}
\noindent\textbf{Queries:} We evaluate the performance of our method and baselines on the queries listed\footnote{We evalaute on more queries, but, to save space, we report only the seven queries in the table; this also explains the weird numbering.} in Table~\ref{tab:query}.  The first five query patterns are derived from the paper~\cite{DBLP:journals/pvldb/MhedhbiS19}, and are used for graph datasets, while the latter are typical sparse tensor kernels~\cite{DBLP:journals/pacmpl/KjolstadKCLA17}. For the triangle query, Q1, we used two variations, Q1d and Q1u, representing triangles on a directed and undirected graph respectively.  We run all other queries only on the directed graphs.

\addvspace{\smallskipamount}
\noindent\textbf{Metrics:} We evaluate the performance of \name and other systems by measuring their overall execution time. Specifically, we track the wall-clock time taken by each system to complete each query from start to finish. This timing excludes the duration spent on data loading, result output, and data statistics collection, but it does account for the time used to create indexes.

We conduct each measurement three times and report the average runtime. 
For Soufflé, we also exclude the compilation time and report only the runtime of the generated programs after compilation. 
For Umbra and Kùzu, we run the query one extra time initially, to avoid the cold start issue, such as additional query compilation or data loading time.  
We set a timeout of 10,000 seconds for each individual experiment repetition.  
Any timeout is marked by X in our graphs.

\addvspace{\smallskipamount}
\noindent\textbf{Environment:} We conduct our experiments on Intel Xeon Phi Server, which has 60 cores and 120 hyperthreads on 4 sockets, as well as 1TB main memory. Except Umbra, \name and the open-source baselines are all compiled using GCC 13.3.0 with Release mode on Rocky Linux 9.4. 

\subsection{Performance Comparison}\label{subsec:exp:perf}

\begin{figure*}[t]
	\includegraphics[width=.95\linewidth, trim={0 0 0 0}, clip]{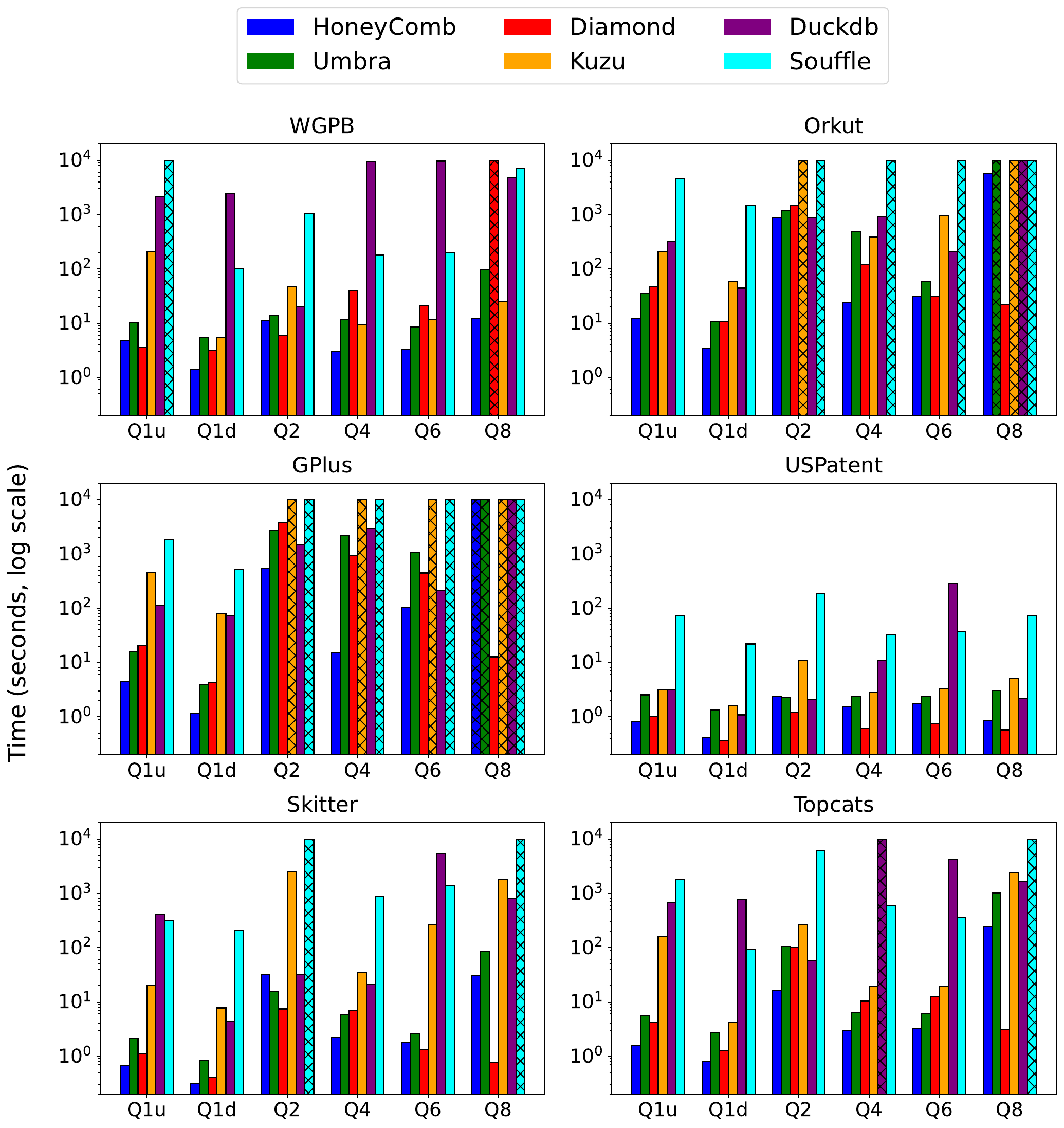}
  \caption{\revB{Performance Comparison on Graph Datasets. An X means ``timeout after 10,000 seconds''}}
  \label{fig:perf}
\end{figure*}

Our first set of experiments compares the runtime of \name and the
other baselines on graph datasets, using all the 60 available
cores. The results,  in Fig.~\ref{fig:perf},  show that \name
consistently outperforms the other systems across all queries and
datasets, with significant runtime reductions observed in most cases,
\revB{with the exception of Diamond, which it outperforms on most but
  not all queries (more on this below).}

On dense datasets, such as GPlus and Orkut, \name benefits from
rewriting and \indexlayout while it efficiently supports lookup and
scan over the sorted dense data. Umbra incurs an overhead during the
construction of the hash trie since most of the sub-tries are touched
and need to be built. \revC{Kùzu, despite using hints for generic join
  order, is slower than Umbra.} DuckDB, although it does not support
the worst-case optimal join and  generally lags behind the others, it performed quite well on the 4-loop query Q2, because Q2 has a small tree-width~\cite{DBLP:conf/pods/KhamisNR16}, where the advantage of the worst-case optimal join over traditional plans diminishes.  Soufflé is the slowest, where almost all queries are timeout, mainly due to the lack of optimizations for both binary and generic joins.

For sparse datasets, such as WGPB, \name continues to outperform the
other systems, but the gap between \name and Umbra
narrows. \revC{Based on the description
  of~\cite{DBLP:journals/pvldb/FreitagBSKN20} we believe that Umbra’s
  lazy trie construction is more effective for sparse data than for
  dense data, because there are fewer contentions between threads and
  fewer sub-trie structures to be constructed on demand.} On sparse
data, \name benefits from the partitioning strategy, which helps
reduce the computational skew. In contrast, both Umbra and Kùzu
partition only on the first attribute and are more likely to suffer
from computation skew. \revB{Diamond behaves slightly better on Q1u
  and Q2 due to its customized operator Expand3 and special
  optimization for queries with small tree-width. However, these
  optimizations are not universally beneficial: Diamond performed
  worse than Umbra on some queries.} DuckDB and Soufflé,
which are not specifically designed for the worst-case optimal join,
perform worse on sparse datasets. Interestingly, Soufflé differs from
DuckDB in that it performs well on queries with large tree-width, such
as the Q6 (4-clique) and Q8 (4-diamond) queries. This is mainly
because Soufflé is optimized for Datalog queries, which are inherently
more cyclic and thus more efficient for these types of queries.

\revB{Umbra and Diamond outperformed \name on several queries on
  USPatent and Skitter: the reason is that the intermediate results on
  these datasets are small, which makes Umbra's lazy trie construction
  and Diamond's hash join very effective.  In contrast, \name computes
  all indices eagerly, and its preprocessing time spent in
  constructing these indices (around 0.5s to 1s) affects more
  significantly the runtime of these relatively cheap queries.

}

{
\begin{table}[t]
	\caption{\revB{Optimizations on Q8 (Orkut)}}
	\label{tab:opt_comp}
	\begin{tabular}{lr}
		\toprule
		Approach & Runtime (s) \\
		\midrule
		\name & 5,680.988 \\
		\name + Tree Decomposition~\cite{DBLP:conf/sigmod/AbergerTOR16} & 1,473.199 \\
		\name + Variable Elimination~\cite{DBLP:conf/pods/KhamisNR16} & 26.956 \\
		\name + Eager Agg~\cite{DBLP:conf/pods/KhamisNR16} & 7.749 \\
		Umbra & Timeout\\
		Diamond (= Umbra + TD) & 436.257 \\
		Diamond + Lookup & 51.513 \\
		Diamond + Loopup + Eager Agg & 21.624 \\
		\bottomrule
	\end{tabular}
\end{table}
}

\revB{For Query Q8, Diamond timed out on WGPB, but outperformed \name
  on all other datasets, due to optimizations that benefit Q8
  specifically.  While \name currently does not support similar
  optimizations, we wanted to get a sense of how much they could
  improve \name's performance.  We hand-optimized Q8 using existing
  optimization techniques, similar to those used in Diamond, and
  present the results in Table~\ref{tab:opt_comp}, using the Orkut
  dataset. Notably, these trends remain consistent across other
  datasets. The results suggest that \name could outperform Diamond
  on queries with good tree decompositions (like Q8), by incorporating
  these orthogonal optimizations.

}

\begin{figure}[t]
	\centering
      \includegraphics[width=0.65\linewidth, trim={0 0 0 0}, clip]{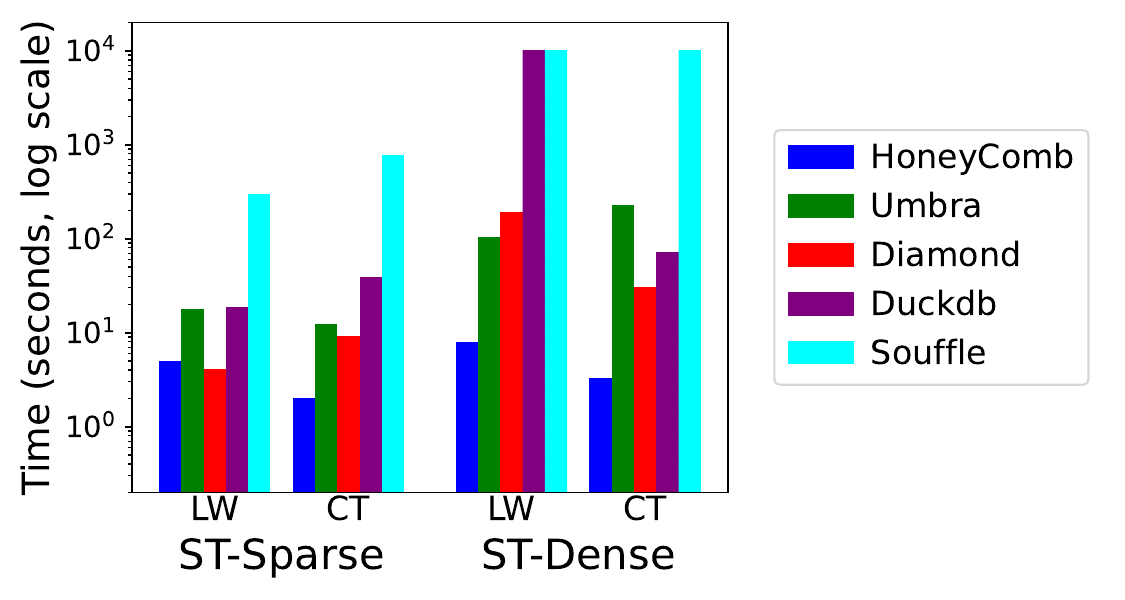}
			\caption{\revB{Performance on Tensor Dataset}}
      \label{fig:perf_tensor}
\end{figure}

Next, we compared \name with the baseline systems on tensor datasets using the queries LW (Loomis-Whitney) and CT (clover-triangle), which have  more complex hypergraphs, typical for tensor kernels. The results, shown in Fig.~\ref{fig:perf_tensor}, demonstrate that \name consistently outperforms most baselines across all queries and datasets, with significant runtime reductions in most cases. We do not include Kùzu, because it doesn't support hypergraph queries. Thus, \name continues to outperform other systems on complex cyclic joins on $n$-ary ($n > 2$) relations.

\subsection{Scalability}\label{subsec:exp:scale}

\begin{figure}[t]
	\centering
	\includegraphics[width=.85\textwidth, trim={0 0 0 0}, clip]{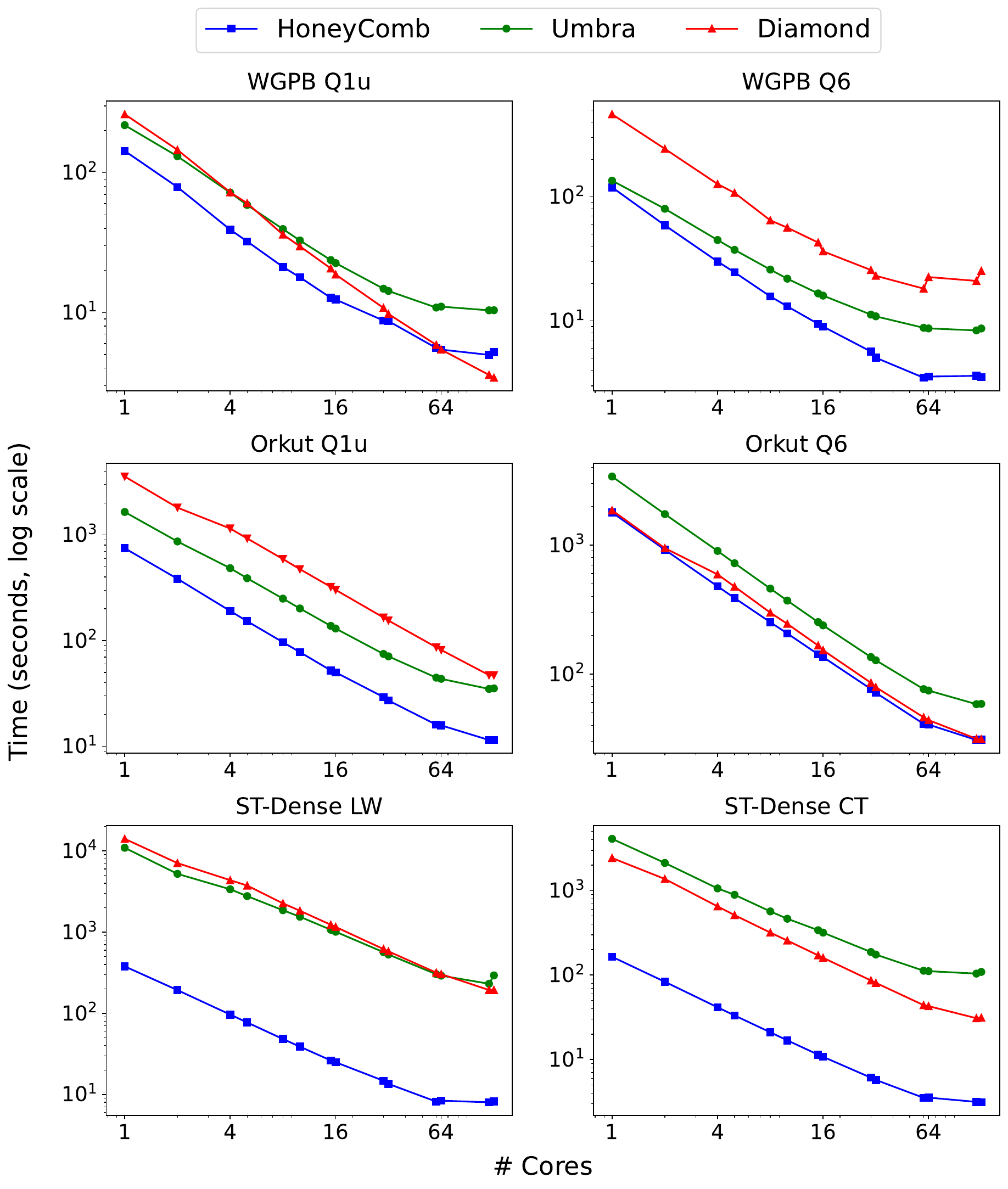}
	\caption{\revC{Speedup of query execution on  $P=1,\ldots,120$
    virtual cores (there are 60 hyperthreaded physical cores).}}
	\label{fig:scale}

\end{figure}

\revC{We evaluated \name's scalability, by measuring the runtime of
  the system as a function of the number of virtual cores $P$.  We report the
  findings only for \name, Umbra and Diamond, since other systems are
  generally too slow to show their scalability. }

We tested the Q1u (triangle) and Q6 (4-clique) queries on the WGPB and Orkut dataset, as well as the LW (Loomis-Whitney) and CT (clover-triangle) queries on the ST-Dense Dataset.

The results are shown in Fig.~\ref{fig:scale}. We find that \name
consistently demonstrated almost linear scalability up to the maximum
number of physical cores, which was 60 on our system.  Increasing the number of
virtual cores up to 120 led to no additional performance improvements at most cases, showing
that the cores were saturated and the use of hyperthreads was not
effective; \revB{we also confirmed that the CPUs were at almost 100\
utilization at 60 virtual cores.}
\revC{In a few cases (Orkut), \name's performance continued to improve a little beyond 60 cores: this is because of a higher contribution to the runtime of exponential search in middle loops, which is less memory efficient. Manually replacing exponential search with linear search restored full CPU utilization at 60 cores.}

Umbra also shows an almost linear speedup, but slightly below that of \name. \revC{For instance, consider Q6 on WGPB, \name is $1.12\times$ faster than Umbra on 1
core, but $2.52\times$ faster on 60 cores, demonstrating that our techniques improve scalability as we increase the number of cores. This can be attributed to several factors.} On sparse datasets, suboptimal partitioning of skewed data can lead to long-tail latencies, negatively impacting performance. On dense datasets, read-write conflicts caused by lazy trie building further hinder speedup when executing queries in parallel, as these conflicts introduce additional overhead that reduces the efficiency of parallel execution.

\revB{Diamond demonstrates similar scalability to Umbra up to 60 cores in most cases. However, Diamond maintains good scalability beyond 60 cores, indicating that hyperthreading benefits its performance. This is mainly due to the use of secondary hash tables over entire relations in Expand3, which are memory-latency bound due to random hash probes over large tables. Hyperthreading, with its separate register stacks and shared execution engine, helps mitigate this bottleneck. Nonetheless, while Diamond achieves speedups beyond 60 cores, its design does not fully address memory access latency, resulting in overall slower performance compared to our system. By leveraging the \indexlayout index and Adaptive Searching, \name achieves better cache locality and effectively utilizes the computational power of cores.}

Also, \name does not achieve optimal speedup on sparse and skew datasets, even though it has better scalability. This is because the partitioning strategy of \name is not only targeted to relieve the skewness problem but also to avoid huge duplicated computation, as mentioned in Sec.~\ref{sec:intro}, which may hurt the scalability. 

Besides, the preprocessing time is not fully parallelized, and the overhead of preprocessing is not negligible in sparse cases.

\revC{These experiments also point to the importance of
  query rewriting for scalability.  Still consider Q6 on WGPB. If we remove
  the query rewriting optimization, then \name is $0.91\times$ faster
  than Umbra (i.e. slower) on 1 core, and $1.15\times$ faster on 60
  cores.  Thus without  rewriting \name gains very little over
  Umbra as $P$ increases (due to its mitigation of skew), but the gain
  becomes more pronounced when we enable the query rewriting optimization.

}

\subsection{Variable Ordering}

\begin{figure}[t]
	\centering
	\begin{minipage}[b]{0.49\textwidth}
		\centering
			  \includegraphics[width=\linewidth, trim={0 0 0 0}, clip]{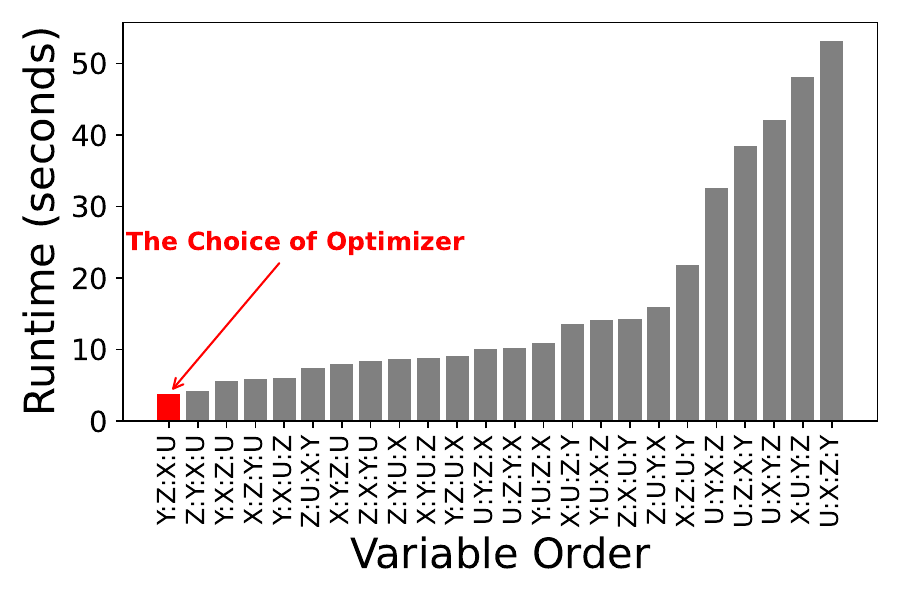}
			  \caption{Effect of Variable Ordering}
			  \label{fig:order}
	\end{minipage}
	~
	\begin{minipage}[b]{0.49\textwidth}
		  \includegraphics[width=\linewidth, trim={0 0 0 0}, clip]{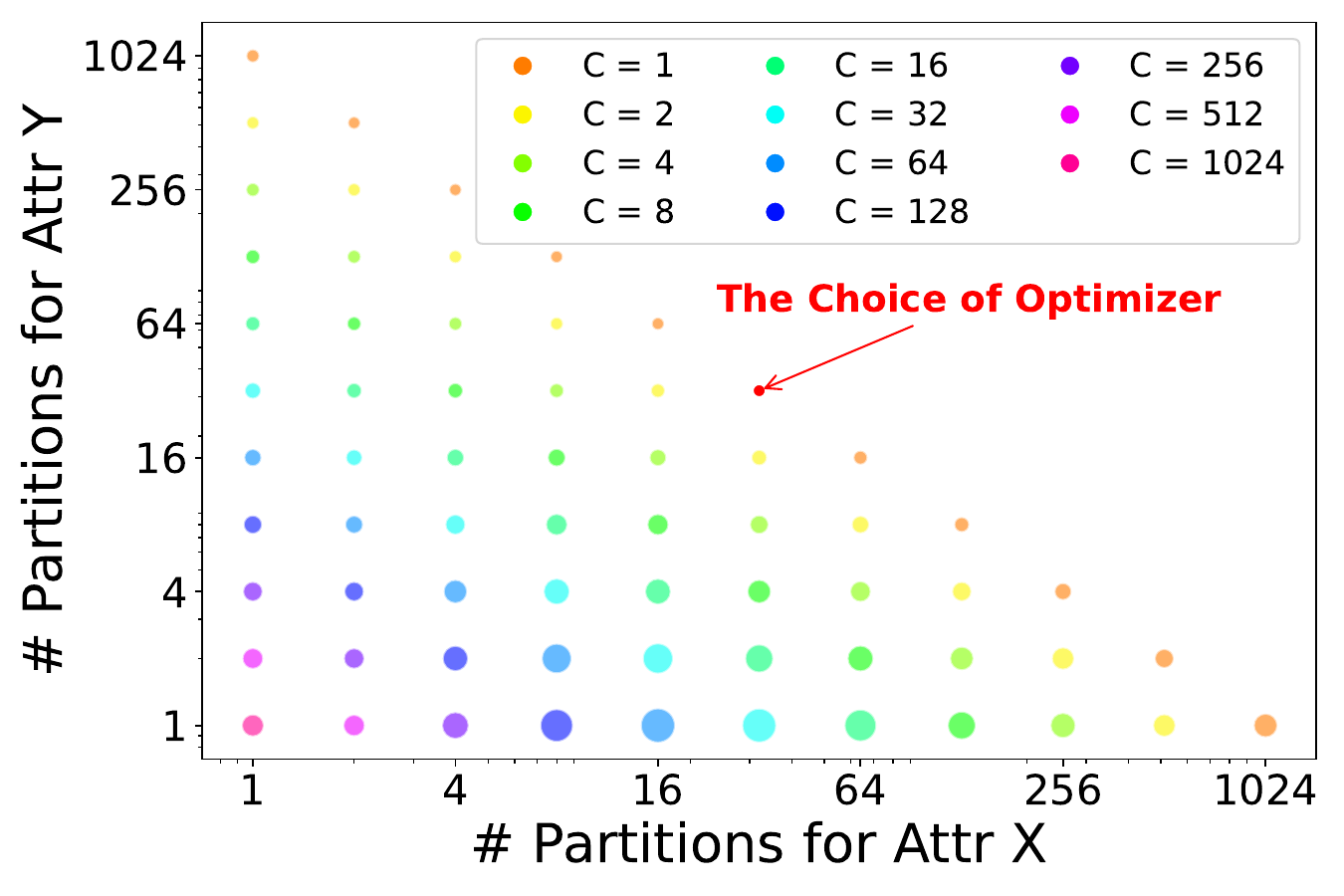}
		  \caption{Effects of Partition Shares}
		  \label{fig:partition}
	\end{minipage}
\end{figure}

Recall that the theoretical analysis of the worst-case optimal join holds for any choice of variable order, however, in practice, the choice of variable order may affect the runtime significantly.  Next, we evaluated how much the variable order can affect the runtime of \name.  

Fig.~\ref{fig:order} shows the runtime for executing the Q6 (4-clique) query on WGPB, for all $24 = 4!$ possible variable orderings.  

The results show that optimizing the variable order can have a significant impact on the query performance. This is because the order in which joins are executed directly influences the size of intermediate results, which can lead to substantial differences in both memory usage and computational effort.

Efficient ordering can minimize the number and length of intersections, reducing the overall complexity of the query execution. In contrast, poor ordering can result in intersecting large sets, which increases memory overhead and slows down execution.

To demonstrate the feasibility of our optimizer, we use a red bar to represent the ordering chosen by our optimizer in Fig.~\ref{fig:order}. In this particular Q6 (4-clique) query, our optimizer chooses the best variable order.  This did not happen for all queries, but it is important to note that the optimizer does not need to select the absolute best ordering, as long as is not significantly slower than the optimal.  

\subsection{Partition}

Next, we conducted a similar analysis on the choice of the partition shares.  We ran the Q1u (triangle) query using $P=1024$ threads, and considered all possible ways to factorize it into $P = P_X \times P_Y \times P_Z$.  The results are shown in Fig.~\ref{fig:partition}: the $X$ and $Y$ axes show $P_X, P_Y$ (the value of $P_Z$ can be inferred, and is also shown by the color of each dot), and the size of each dot represents the runtime (smaller is better).  The optimal shares are $32 \times 32 \times 1$, and these are, indeed, the shares chosen by our optimizer.  The traditional parallelization method, which allocates all shares to one variable, corresponds to the factorization $1024 \times 1 \times 1$, and is the right-most orange dot, which is visibly worse than the optimal choice of shares.  

Recall that the HyperCube algorithm tries to minimize the communication cost; when all three relations of the query have the same size, then the optimal shares are equal, which, in our case corresponds to $8 \times 8 \times 16$ or some permutation thereof.  As one can see from the graph, these choices of shares are far from optimal: our optimizer adjusts the partition shares based on not only the data distribution, but also on query characteristics that we care about. As noted in Sec.~\ref{subsec:plan}, the optimizer will rule out these partition shares by detecting the heavy duplicated intersection computations across different threads.

On the other hand, if we keep $P_Z =1$ then it doesn't matter too much how many shares we allocate to $X$ or $Y$ (the line of red dots, which also contains the optimal configuration).  This is because the data itself is relatively evenly distributed, and the difference in allocating shares between X and Y has little impact on performance.

Overall, the graph shows a large variation in the total runtime, proving the need for an optimized allocation of shares.

\subsection{Query Rewriting and Indexes}

\begin{figure}[t]
	\centering
		  \includegraphics[width=.5\linewidth, trim={0 0 0 0}, clip]{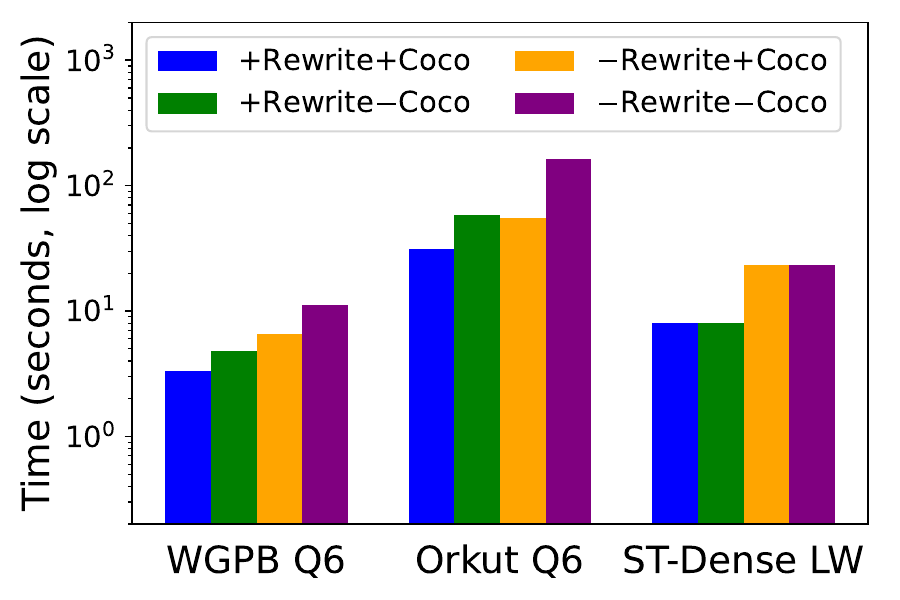}
		  \caption{\revC{Effects of Rewriting and \indexlayout}}
		  \label{fig:other}
\end{figure}

In this section, we present additional experimental results that highlight key aspects of our optimization techniques.

We first evaluate the effectiveness of our query rewriting optimization by measuring the runtime with and without rewriting enabled. The results, summarized in Fig.~\ref{fig:other}, show that enabling query rewriting consistently improves performance across all queries. This demonstrates that our rewriting strategies optimize execution plans, leading to more efficient query processing. \revC{For the the queries LW on the ST-Dense dataset, the rewriting is not effective, due to no duplicated intersections in its plan.}

Next, we assess the impact of our \indexlayout index on query performance. We compare the runtime with the \indexlayout index enabled and disabled, and the results indicate a significant performance improvement when the index is utilized. Also, we report the size of the \indexlayout index, which ranges from $1$GB to $5$GB for each table and is comparable with datasets. Thus, the \indexlayout does not introduce significant storage overhead. This ensures that the index remains practical for optimizing joins without imposing a large memory footprint.

In addition, we measure the total time spent in the preprocessing stage, which includes the creation of the \indexlayout index. The preprocessing time ranges from $0.2$s to $5$s, depending on the number of tables and the size of datasets. For long-running queries, the preprocessing time is generally minimal compared to the join execution time. For short-running queries, even sometimes the preprocessing time is larger than the join time, but the total runtime is still less than or comparable to other baselines. This indicates that the overhead introduced by preprocessing is acceptable, making our approach efficient and suitable for real-time query processing.

\section{Conclusion}\label{sec:conclusion}

We introduced \name, an implementation of Generic Join (a special case
of Worst Case Optimal Join) on a multicore, shared-memory
architecture.  Prior parallel implementations of WCOJ partitioned only
the top loop variable.  Instead, in \name we partition the domains of
all variables, which reduces the computation skew.  We also
introduced a simple trie index, based on a novel two-stage
sorting-based parallel algorithm, which can be constructed eagerly and
cheaply, and removes concurrent conflicts during the query evaluation.
The use of a sorted array instead of a hash table also takes advantage
of modern hardware by improving cache locality and enabling vector
processing.  We co-optimized the choice of the variable order and the
computation of the optimal shares by using a novel cost model for data
skew, and described a rewriting technique for WCOJ that reduced the
amount of redundant computations.  Finally, we reported an extensive
evaluation of our implementation, proving the effectiveness of the
partitioning strategy, of the optimized choice of variable order and
shares, and of the rewriting technique.

Future work includes further leveraging hardware by supporting vectorization, 
speeding up the optimization time through dynamic programming, and extending
the rewriting technique to support more complicated cases.

\newpage

\bibliographystyle{ACM-Reference-Format}
\bibliography{wcoj}

\end{document}